\documentclass[a4paper,aps,preprintnumbers,showpacs,twocolumn,superscriptaddress,nofootinbib]{revtex4-1}

\usepackage{amsmath}
\usepackage{amssymb}
\usepackage{graphicx}
\usepackage{epsfig}
\usepackage{color}
\usepackage{url}
\usepackage{times}
\usepackage{bm}         

\newcommand{\cf}{cf.~}
\newcommand{\ie}{i.e.,~}

\newcommand{\be}{\begin{equation}}
\newcommand{\ee}{\end{equation}}
\newcommand{\bea}{\begin{eqnarray}}
\newcommand{\eea}{\end{eqnarray}}
\newcommand{\nn}{\nonumber}
\newcommand{\noi}{\noindent}
\newcommand{\bwt}{\begin{widetext}}
\newcommand{\ewt}{\end{widetext}}

\font\tenscr=rsfs10 scaled1100
\font\sevenscr=rsfs7 
\font\fivescr=rsfs5 
\skewchar\tenscr='177
\skewchar\sevenscr='177
\skewchar\fivescr='177
\newfam\scrfam
\textfont\scrfam=\tenscr
\scriptfont\scrfam=\sevenscr
\scriptscriptfont\scrfam=\fivescr

\def\scri{{\fam\scrfam I}}
\def\scre{{\fam\scrfam E}}

\begin{document}

\title{Black-hole horizons as probes of black-hole dynamics II: geometrical
  insights}

\author{Jos\'e Luis Jaramillo}
\affiliation{
Max-Planck-Institut f{\"u}r Gravitationsphysik, Albert Einstein
Institut, Potsdam, Germany 
}

\author{Rodrigo P. Macedo}
\affiliation{
Max-Planck-Institut f{\"u}r Gravitationsphysik, Albert Einstein
Institut, Potsdam, Germany 
}
\affiliation{
Theoretisch-Physikalisches Institut, Friedrich-Schiller-Universit{\"a}t Jena, Jena, Germany
}

\author{Philipp Moesta}
\affiliation{
Max-Planck-Institut f{\"u}r Gravitationsphysik, Albert Einstein
Institut, Potsdam, Germany 
}

\author{Luciano Rezzolla}
\affiliation{
Max-Planck-Institut f{\"u}r Gravitationsphysik, Albert Einstein
Institut, Potsdam, Germany 
}
\affiliation{
Department of Physics and Astronomy,
Louisiana State University,
Baton~Rouge, Louisiana, USA
}

\begin{abstract}
In a companion paper~\cite{Jaramillo:2011re}, we have presented a
cross-correlation approach to near-horizon physics in which bulk
dynamics is probed through the correlation of quantities defined at
inner and outer spacetime hypersurfaces acting as test screens. More
specifically, dynamical horizons provide appropriate inner screens in
a $3+1$ setting and, in this context, we have shown that an
effective-curvature vector measured at the common horizon produced in
a head-on collision merger can be correlated with the flux of linear
Bondi-momentum at null infinity. In this paper we provide a more sound
geometric basis to this picture. First, we show that a
\textit{rigidity} property of dynamical horizons, namely foliation
uniqueness, leads to a preferred class of null tetrads and Weyl
scalars on these hypersurfaces. Second, we identify a heuristic
horizon newslike function, depending only on the geometry of spatial
sections of the horizon. Fluxes constructed from this function offer
refined geometric quantities to be correlated with Bondi fluxes at
infinity, as well as a contact with the discussion of quasilocal
4-momentum on dynamical horizons. Third, we highlight the importance
of tracking the internal horizon dual to the apparent horizon in
spatial 3-slices when integrating fluxes along the horizon. Finally,
we discuss the link between the dissipation of the nonstationary part
of the horizon's geometry with the viscous-fluid analogy for black
holes, introducing a geometric prescription for a ``slowness
parameter'' in black-hole recoil dynamics.
\end{abstract}

\maketitle

\section{Introduction}
\label{s:geom_Hor}

In Ref.~\cite{Jaramillo:2011re} (paper I hereafter) a {\em
  cross-correlation} methodology for studying near-horizon
strong-field physics was outlined. Spacetime dynamics was probed
through the cross-correlation of timeseries $h_{\mathrm{inn}}$ and
$h_{\mathrm{out}}$ defined as geometric quantities on {\em inner} and
{\em outer} hypersurfaces, respectively. The latter are understood as
{\em test screens} whose geometries respond to the bulk dynamics, so
that the (global) functional structure of the constructed
cross-correlations encodes some of the features of the bulk geometry.
This is in the spirit of reconstructing spacetime dynamics in an {\em
  inverse-scattering} picture. In the context of asymptotically flat
black-hole (BH) spacetimes, the BH event horizon $\scre$ and future
null infinity $\scri^+$ provide natural test hypersurfaces from a
global perspective. However, when a $3+1$ approach is adopted for the
numerical construction of the spacetime, dynamical trapping horizons
${\cal H}$ provide more appropriate hypersurfaces to act as inner test
screens\footnote{In paper I future outer trapping horizons were
  denoted by ${\cal H}^+$ to distinguish them from past outer trapping
  horizons ${\cal H}^-$ occurring in the Robinson-Trautman model,
  extending the study in~\cite{Rezzolla:2010df}.}. In the application
of this correlation strategy to the study of BH post-merger recoil
dynamics, an effective-curvature vector $\tilde{K}^\mathrm{eff}_i(v)$
was constructed~\cite{Jaramillo:2011re} on ${\cal H}$ as the quantity
$h_{\mathrm{inn}}$ to be cross-correlated with $h_{\mathrm{out}}$,
where the latter is the flux of Bondi linear momentum
$(dP_i^{\mathrm{B}}/du)(u)$ at $\scri^+$ (here, $u$ and $v$ denote,
respectively, {advanced} and {retarded
  times\footnote{Cross-correlation of quantities at ${\cal H}$ and
    $\scri^+$ requires the choice of a gauge mapping between the
    advanced and retarded times $u$ and $v$. This time-stretching
    issue is discussed in paper I.}). 
In this paper we explore some
  geometric structures underlying and extending the heuristic
  construction in~\cite{Jaramillo:2011re} of this
  effective local probe into BH recoil dynamics.
 
The adaptation of geometric structures and tools from $\scri^+$ to BH
horizons is at the basis of important geometric developments in BH
studies, notably the quasilocal frameworks of isolated and dynamical
trapping horizons~\cite{Hayward94a,Ashtekar03a,Ashtekar:2004cn} (see
also Refs.~\cite{Haywa94c,Haywa03}). In this spirit, the construction
of $\tilde{K}^\mathrm{eff}_i(v)$ on the horizon ${\cal H}$ partially
mimics the functional structure of the flux of Bondi linear momentum
at $\scri^+$. In particular, $(dP_i^{\mathrm{B}}/du)(u)$ can be
expressed in terms of (the dipolar part of) the square of the news
function ${\cal N}$ on sections of $\scri^+$, whereas the definition
of $\tilde{K}^\mathrm{eff}_i(v)$ involves the (dipolar part of the)
square of a function $\tilde{\cal N}$ constructed from the Ricci
scalar ${}^2\!R$ on sections of ${\cal H}$. However, the functions
${\cal N}$ and $\tilde{\cal N}$ differ in their spin-weight and, more
importantly, they show a different behavior in time: whereas ${\cal
  N}(u)$ is an object well-defined in terms of geometric quantities on
time sections ${\cal S}_u\subset {\scri^+}$, nothing guarantees this
{\em local-in-time} character of $\tilde{\cal N}(v)$ [see
  Eq.~(\ref{e:news_R_v}) below]. The latter is a crucial
characteristic of the news function, so that $\tilde{\cal N}(v)$
cannot be considered as a valid {\em newslike} function on ${\cal
  H}$.

These structural differences suggest that, in spite of the success of
$\tilde{K}^\mathrm{eff}_i$ in capturing effectively (at the horizon)
some qualitative aspects of the flux of Bondi linear momentum (at null
infinity), a deeper geometric insight into the dynamics of ${\mathcal
  H}$ can provide hints for a {\em refined} correlation treatment. In
this context, the specific goals in this paper are: i) to justify the
role of $\tilde{K}^\mathrm{eff}_i$ as an {\em effective} quantity to
be correlated to $(dP_i^{\mathrm B}/du)$, suggesting candidates
offering a refined version; ii) to explore the introduction of a valid
newslike function on ${\cal H}$, only depending on the geometry of
sections ${\cal S}_v\subset{\cal H}$; iii) to establish a link between
the cross-correlation approach in~\cite{Jaramillo:2011re} and other
approaches to the study of the BH recoil based on quasilocal
momentum.

The paper is organized as follows. Section~\ref{s:evol_system}
introduces the basic elements on the inner screen ${\cal H}$ geometry
and revisits the effective-curvature vector of paper I. Aiming at
understanding the dynamics of the latter, a geometric system governing
the evolution of the intrinsic curvature along the horizon ${\cal H}$
is discussed, making apparent the key driving role of the Weyl tensor.
In Sec.~\ref{s:fundamental_DH} some fundamental results on
dynamical horizons are discussed, in particular a {\em rigidity
  structure} enabling a preferred choice of null tetrad on ${\cal H}$.
Proper contractions of the latter with the Weyl tensor lead in
Sec.~\ref{s:Bondi_DH_linearmomentum} to newslike functions and
associated {\em Bondi-like fluxes} on ${\cal H}$ providing refined
quantities on the horizon to be correlated with Bondi fluxes at
$\scri^+$, as well as making contact with quasilocal approaches to BH
linear momentum. In Sec.~\ref{s:viscous_picture} our
geometric discussion is related to the viscous-fluid analogy of BH
horizons, providing in particular a geometric prescription for the
{\em slowness parameter} $P$ in~\cite{Price:2011fm}. Conclusions are
presented in Sec.~\ref{s:conclusions}. Finally a first appendix 
gathers the geometric notions in the text, whereas a second appendix
emphasizes the physical relevance of {\em internal horizons} when computing
fluxes along ${\cal H}$. We use a spacetime
signature $(-,+,+,+)$, with abstract index notation (first letters,
$a$, $b$, $c$..., in Latin alphabet) and Latin midalphabet indices,
$i,j,k...$, for spacelike vectors. We also employ the standard
convention for the summation over repeated indices. All the quantities
are expressed in a system of units in which $c=G=1$.

\section{Geometric evolution system  
on the horizon: the role of the Weyl tensor}
\label{s:evol_system}

\subsection{The inner screen ${\cal H}$}
\label{s:inner_screen}

Let us consider a BH spacetime $({\cal M}, g_{ab})$, with associated
Levi-Civita connection $\nabla_a$, endowed with a $3+1$ spacelike
foliation $\{\Sigma_t\}$. Let us consider an {\em inner} hypersurface
${\cal H}$, to be later identified with the BH horizon, such that the
intersection of the slices $\Sigma_t$ with the world-tube ${\cal H}$
defines the foliation of ${\cal H}$ by closed spacelike surfaces
$\{{\cal S}_t\}$. We consider an evolution vector $h^a$ along ${\cal
  H}$, characterized as that vector tangent to ${\cal H}$ and normal
to the slices $\{{\cal S}_t\}$ that transports the slice ${\cal S}_t$
onto the slice ${\cal S}_{t+\delta t}$. The normal plane at each
point of ${\cal S}_t$ can be spanned in terms of the {\em outgoing}
null vector $\ell^a$ and the {\em ingoing} vector $k^a$, chosen to
satisfy $\ell^a k_a=-1$. Directions of $\ell^a$ and $k^a$ are fixed,
though a rescaling freedom remains (see Fig.\ref{fig:vectors}). In particular, and without loss of
generality in our context, we can write~\cite{Booth04a}
\begin{eqnarray}
\label{e:vector_h}
h^a = \ell^a - C k^a  ,
\end{eqnarray}
so that $h^ah_a = 2C$. Therefore: $h^a$ is, respectively, 
spacelike if $C>0$, null if $C=0$, and timelike if $C<0$.

\begin{figure}[t!]
\begin{center}
\includegraphics[angle=-90,width=8.0cm,clip=true]{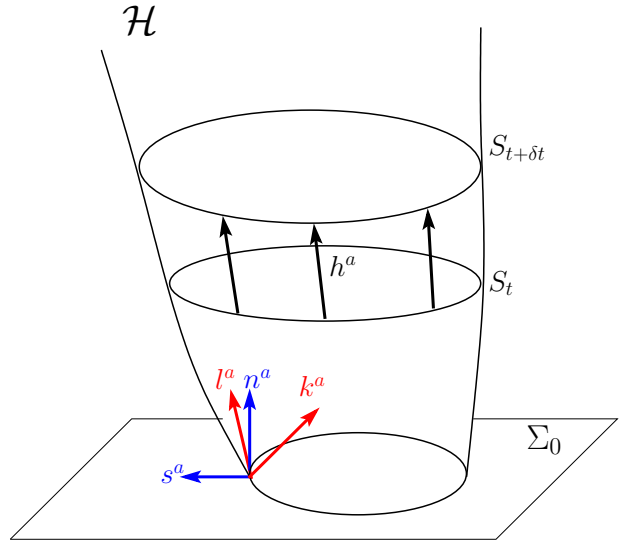}
\end{center}
\caption{ Worldtube ${\cal H}$ foliated by closed spacelike surfaces
  $\{{\cal S}_t\}$ as the result of a $3+1$ spacelike foliation
  $\{\Sigma_t\}$. The evolution vector $h^a$ (tangent to ${\cal H}$
  and normal to $\{{\cal S}_t\}$) transports the slice ${\cal S}_t$ to
  ${\cal S}_{t+\delta t}$. The normal plane at each
  point of ${\cal S}_t$ can be spanned by the outgoing and ingoing
null normal vectors $\ell^a$ and $k^a$ or by $n^a$, the unit timelike normal to
  $\Sigma_t$, and $s^a$, the spacelike outgoing normal to $S_t$ and
  tangent to $\Sigma_t$ (\cf Appendix \ref{a:geometry_Hor}) . 
  } \vglue-0.5cm
\label{fig:vectors}
\end{figure}

Regarding the intrinsic geometry on ${\cal S}_t$, the induced metric
is denoted by $q_{ab}$, its Levi-Civita connection by ${}^2\!D_a$ and
the corresponding Ricci curvature scalar by ${}^2\!R$.  The area form
is ${}^2\!\epsilon = \sqrt{q} dx^1\wedge dx^2$ and we will denote the
area measure as $dA=\sqrt{q}d^2x$.  The infinitesimal evolution of the
intrinsic geometry along ${\cal H}$, \ie the evolution of the induced
geometry $q_{ab}$ along $h^a$, defines the {\em deformation tensor}
$\Theta^{(h)}_{ab}$ [\cf Equation~(\ref{e:deformation_tensor}) in Appendix
\ref{a:geometry_Hor}]
\begin{eqnarray}
\label{e:deformation_tensor_h}
\Theta^{(h)}_{ab} \equiv \frac{1}{2} 
 {\delta}_h q_{ab} = \sigma^{(h)}_{ab} +  \frac{1}{2} \theta^{(h)} q_{ab} \,,
\end{eqnarray}
where the trace $\theta^{(h)}=\Theta^{(h)}_{ab}q^{ab}$, referred to as
the {\em expansion} along $h^a$, measures the infinitesimal evolution
of the ${\cal S}_t$ element of area along ${\cal H}$, whereas the
traceless {\em shear} $\sigma^{(h)}_{ab}$ controls the deformations of
the induced metric (see Eq.~(\ref{e:expansion_shear}) in Appendix
\ref{a:geometry_Hor}).  Here $\delta_h$ can be identified with the
projection on ${\cal H}$ of the Lie derivative ${\cal L}_h$ [see
  Eq.~(\ref{e:delta_v_X}) and the remark after Eq.~(\ref{e:kappa_v})].
Before reviewing the effective-curvature vector
$\tilde{K}^\mathrm{eff}_i$, let us discuss the time parametrization of
${\cal H}$.

We recall that jumps of apparent horizons (AHs) are generic in $3+1$
evolutions of BH spacetimes. The dynamical trapping horizon framework
offers a spacetime insight into this behavior by understanding the
jumps as corresponding to marginally trapped sections of a (single)
hypersurface bending in spacetime, but multiply foliated by spatial
hypersurfaces in the $3+1$ foliation $\{\Sigma_t\}$
\cite{Booth:2005ng, Nielsen:2005af, Schnetter-Krishnan-Beyer-2006,
  Booth:2007wu, Jaramillo:2009zz}. In the particular case of binary BH
mergers this picture predicts, after the moment of its first
appearance, the splitting of the common AH into two horizons: a
growing {\em external} common horizon and a shrinking {\em internal}
common horizon~\cite{Schnetter-Krishnan-Beyer-2006,
  Jaramillo:2009zz}. It is standard to track the evolution of the
external common horizon, the proper AH, but to regard the internal
common horizon as physically irrelevant. In Appendix~\ref{appendixB}
we stress however the relevance of the internal horizon in the context
of the calculation of physical fluxes into the black-hole singularity.

In Fig.~\ref{fig:advanced_time} we illustrate this picture in a
simplified (spherically symmetric) collapse scenario that retains the
relevant features of the discussion. On one side, the relevant outer
screen boundary (namely, null infinity $\scri^+$) is parametrized by
the retarded time $u$, something explicitly employed in the expression
of the flux of Bondi momentum in Eqs. (33) and (34) of paper I. On the
other side, from the $3+1$ perspective, the moment $t_c$ of first
appearance of the (common) horizon corresponds to the coordinate time
$t$ at which the $3+1$ foliation $\{\Sigma_t\}$ firstly intersects the
dynamical horizon ${\cal H}$. For $t>t_c$, $\Sigma_t$ slices intersect
twice (multiply, in the generic case) the hypersurface ${\cal H}$
giving rise to the external and internal common horizons (\cf ${\cal
  H}$ in Fig.~\ref{fig:advanced_time}). Therefore, the time function
$t$ is not a good parameter for the whole dynamical horizon ${\cal
  H}$. An appropriate parametrization of this hypersurface ${\cal H}$
is given in terms of an advanced time, such as $v$, parametrizing past
null infinity $\scri^-$. More precisely, (for a spacelike world-tube
portion of ${\cal H}$) we can label sections of ${\cal H}$ by an
advanced time $v$ starting from an initial value $v_0$ corresponding
to the first $v= \mathrm{const}$ null hypersurface hitting the
spacetime singularity, \ie ${\cal H}= \bigcup_{v\geq v_0} {\cal S}_v$.

\begin{figure}[t!]
\begin{center}
\includegraphics[width=6.0cm,clip=true]{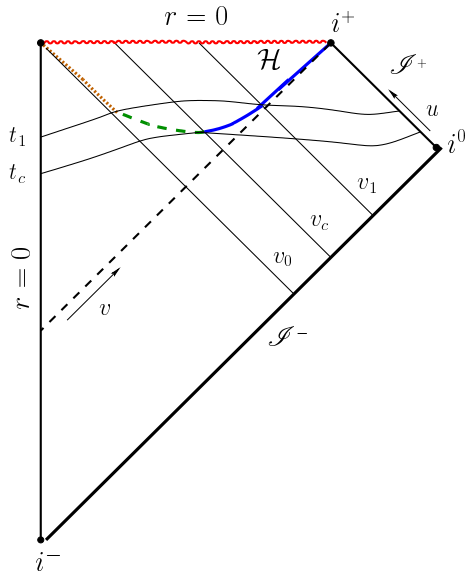}
\end{center}
\caption{ Carter-Penrose diagram (corresponding, for simplicity, to a
  generic spherically symmetric collapse) illustrating the time
  parametrization of the outer and inner screens. The outer boundary
  given by $\scri^+$ is properly parametrized by the retarded time
  $u$, whereas an advanced time $v$ runs along inner boundaries, in
  particular the dynamical horizon ${\cal H}$. Given a $3+1$ foliation
  $\{\Sigma_t\}$, $t_c$ denotes the time $t$ at which the horizon
  first appears. For $t>t_c$, $\Sigma_t$ slices intersect multiply the
  hypersurface ${\cal H}$, giving rise to internal and external
  horizons.  On the contrary, the advanced coordinate $v$ provides a
  good parametrization of ${\cal H}$ from an initial $v>v_0$.  }
\label{fig:advanced_time}
\end{figure}

\subsection{Effective-curvature vector $\tilde{K}^{\mathrm{eff}}_i$}

In paper I the effective-curvature vector $\tilde{K}^\mathrm{eff}_i$
was introduced using the parametrization of ${\cal H}$ by the time
function $t$ associated with the spacetime $3+1$ slicing. In
particular, $\tilde{K}^\mathrm{eff}_i(t)$ was defined only on the
external part of the horizon ${\cal H}$, for $t\geq t_c$.  We can now
extend the definition of $\tilde{K}^\mathrm{eff}_i$ to the whole
horizon ${\cal H}$ (more precisely, to a spacelike world-tube portion
of it) by making use of its parametrization by the advanced time $v$
adapted to the $3+1$ slicing of ${\cal H}$.  Given a section ${\cal
  S}_v\subset {\cal H}$, we consider a vector $\xi^i$ transverse to it
(\ie generically not tangent to ${\cal S}_v$) and tangent to the
3-slice $\Sigma_t$ that intersects ${\cal H}$ at ${\cal S}_v$ (\ie
${\cal S}_v={\cal H}\cap\Sigma_t$).  Then, the component
$\tilde{K}^\mathrm{eff}[\xi](v)$ is expressed as\footnote{For avoiding
  the introduction of lapse functions related to different
  parametrizations of ${\cal H}$, we postpone the fixing of the
  coefficient to Sec.~\ref{s:Bondi_DH_linearmomentum}. We note that
  a global constant factor is irrelevant for cross-correlations. }
\begin{eqnarray}
\label{e:3D_Keff_v}
\tilde{K}^\mathrm{eff}[\xi](v) 
\propto  -\oint_{{\cal S}_v} (\xi^is_i) \left(\tilde{\cal N}(v)\right)^2 dA \,,
\end{eqnarray}
where $s_i$ is the spacelike normal to ${\cal S}_v$ and tangent
to $\Sigma_t$, and  
\begin{eqnarray}
\label{e:news_R_v}
\tilde{\cal N}(v) 
\equiv \int_{v_0}^v {}^2\!R(v') dv' + \tilde{\cal N}_{v_0}\,,
\end{eqnarray}
where $ {}^2\!R$ is the Ricci curvature scalar on $({\cal S}_v,
q_{ab})$ and $\tilde{\cal N}_{v_0}$ is an initial function to be
fixed.  As commented above, in spite of the formal similarity with the
news function ${\cal N}(u)$ at $\scri^+$ [\cf Equation~(34) in paper I),
definition (\ref{e:news_R_v}] does not guarantee the local-in-time
character of $\tilde{\cal N}(v)$ since it is expressed in terms of a
time integral on the past history.

In order to study the dynamics of $\tilde{K}^\mathrm{eff}_i$, we
consider the evolution of the Ricci scalar curvature ${}^2\!R$ along
the world-tube ${\cal H}$ . In terms of the elements introduced above,
the evolution of the Ricci scalar curvature ${}^2\!R$ along $h^a$ has
the form
\be
\label{e:R_evolution}
\delta_h {}^2\!R = - \theta^{(h)}\,{}^2\!R 
+ 2 \; {}^2\!D^a{}^2\!D^b \sigma^{(h)}_{ab}
- {}^2\!\Delta \theta^{(h)} \ \,,
\ee 
where ${}^2\!\Delta=q^{ab}{}^2\!D_a{}^2\!D_b $ denotes the Laplacian
on ${\cal S}_t$. Expression (\ref{e:R_evolution}) is a fundamental one
in our work and it applies to \textit{any} hypersurface ${\cal H}$
foliated by closed surfaces ${\cal S}_t$.  Contact with BHs is made
when ${\cal H}$ is taken as the spacetime {event horizon} or as the
{dynamical horizon} associated with the foliation $\{\Sigma_t\}$.

\subsection{Geometry evolution on BH horizons}
\label{s:BHs}

We briefly recall the notions of BH horizon relevant here and refer to
Appendix~\ref{a:geometry_Hor} for a systematic presentation of the
notation. First, the event horizon (EH) $\scre$ is the boundary of the
spacetime region from which no signal can be sent to $\scri^+$, \ie
the region in ${\cal M}$ not contained in the causal past
$J^-(\scri^+)$ of $\scri^+$. The EH is a null hypersurface,
characterized as $\scre = \partial J^-(\scri^+)\cap {\cal M}$. Second,
a dynamical horizon (DH) or (dynamical) {\em future outer trapping
  horizon} ${\cal H}$ is a quasilocal model for the BH horizon
based on the notion of a world-tube of AHs. More specifically, a future
 outer trapping  horizon ${\cal H}$ is a hypersurface that can 
 be foliated by marginally
(outer) trapped surfaces ${\cal S}_t$, \ie ${\cal H} =
\bigcup_{t\in\mathbb{R}} {\cal S}_t$ with outgoing expansion
$\theta^{(\ell)}= 0$ on ${\cal S}_t$, satisfying: i) a {\em future}
condition $\theta^{(k)}<0$, and ii) an {\em outer condition}
$\delta_k\theta^{(\ell)}<0$. In the dynamical regime, \ie when matter
and/or radiation cross the horizon (namely when
$\delta_\ell\theta^{(\ell)}\neq 0$), the outer condition is equivalent
to the condition that ${\cal H}$ is
spacelike~\cite{Booth:2006bn}\footnote{This property actually
  substitutes the outer condition in the DH
  characterization~\cite{Ashtekar02a,Ashtekar03a} of quasilocal
  horizons.}. Therefore, for dynamical trapping horizons we have $C>0$
in Eq.  (\ref{e:vector_h}) [\cf discussion after
  Eq.~(\ref{e:TH_condition})].

For both EHs and DHs, an important {area theorem} holds:
\mbox{$\delta_h A = \oint_{{\cal S}_t} \theta^{(h)} dA >0$}. In the
case of an EH, Hawking's area theorem~\cite{Hawking71a,Hawking72a}
guarantees the growth of the area, whereas in the case of a DH, the
positivity of $\delta_h A = -\oint_{{\cal S}_t} C \theta^{(k)} dA$
[\cf Equation~\ref{e:theta_sigma_h}] is guaranteed by its spacelike
character ($C>0$) together with the future condition $\theta^{(k)}<0$.

We make now contact with Eq.~(\ref{e:R_evolution}) and interpret the
elements that determine the dynamics of ${}^2\!R$. The growth of the
area of a BH horizon guarantees the (average) positivity of
$\theta^{(h)}$. This offers a qualitative understanding of the
dynamical decay of ${}^2\!R$: the first term in the right-hand side
drives an exponential-like decay of the Ricci scalar curvature. More
precisely, nonequilibrium deformations of the Ricci scalar curvature
${}^2\!R$ in BH horizons decay exponentially as long as the horizon
grows in area.  Regarding the elliptic operators acting on the shear
and the expansion [second and third terms in the right hand side of
  Eq.~(\ref{e:R_evolution})] they provide dissipative terms smoothing
the evolution of ${}^2\!R$. Indeed, in Sec.~\ref{s:viscous_picture}
we will review a viscosity interpretation of $\theta^{(h)}$ and
$\sigma^{(h)}_{ab}$, in particular associating with them respective
decay and oscillation timescales of the horizon geometry.

\subsubsection{Complete evolution system driving ${}^2\!R$}

A further understanding of Eq.~(\ref{e:R_evolution}) requires a
control of the dynamics of the shear $\sigma^{(h)}_{ab}$, of the
expansion $\theta^{(h)}$ and of the induced metric $q_{ab}$, the
latter controlling the elliptic operators ${}^2\!D^a{}^2\!D^b$ and
${}^2\!\Delta$. Therefore, we need evolution equations determining
$\delta_h q_{ab}$, $\delta_h\theta^{(h)}$ and
$\delta_h\sigma^{(h)}_{ab}$:

\smallskip
i) $\delta_h q_{ab}$: {\em definition of the deformation tensor}. The
evolution of $q_{ab}$ is dictated by $\sigma^{(h)}_{ab}$ and
$\theta^{(h)}$ [\cf Equation~(\ref{e:deformation_tensor_h})].

\smallskip
ii) $\delta_h\theta^{(h)}$: {\em focusing or Raychadhuri-like
  equation}. The evolution of $\theta^{(h)}$ involves the Ricci
tensor $R_{ab}$, \ie the ``trace part'' of the spacetime Riemann
tensor ${R^a}_{bcd}$, thus introducing the stress-energy tensor
$T_{ab}$ through Einstein equations.

\smallskip
iii) $\delta_h\sigma^{(h)}_{ab}$: {\em tidal equation}. The evolution
of $\sigma^{(h)}_{ab}$ is driven by the Weyl tensor ${C^a}_{bcd}$,
\ie the traceless part of the spacetime Riemann tensor, thus
involving dynamical gravitational degrees of freedom but not directly
the Einstein equations.

\medskip
The structural feature that we want to underline about these equations
is shared by evolution systems on EHs and DHs, although the explicit
form of the equations differ in both cases.  More specifically,
whereas for EHs the evolution equations for ${}^2\!R$, $q_{ab}$,
$\theta^{(h)}$ and $\sigma^{(h)}_{ab}$ form a ``closed'' evolution
system, in the DH case additional geometric objects (requiring further
evolution equations) are brought about through the evolution equations
$\delta_h q_{ab}$, $\delta_h\theta^{(h)}$ and $\delta_h
\sigma^{(h)}_{ab}$. Moreover, an explicit dependence on the function
$C$, related to the choice of $3+1$ slicing as discussed later [\cf
  Equation~(\ref{e:condition_C})], is involved in the DH case.  For these
reasons, and for simplicity, in the rest of this subsection we
restrict our discussion to the case of an EH, indicating that the main
qualitative conclusion also holds for DHs, whose details will be
addressed elsewhere.

The EH $\scre$ is a null hypersurface generated by the
evolution vector $h^a$, a null vector in this case: $h^a= \ell^a$.
The null generator $\ell^a$ satisfies a pregeodesic equation
$\ell^c\nabla_c \ell^a = \kappa^{(\ell)} \ell^a$ [see
  Eq.~(\ref{e:kappa_v}) for the expression of the nonaffinity
  parameter $\kappa^{(\ell)}$]. Choosing an affine reparametrization
such that $\ell^a$ is geodesic, \ie $\kappa^{(\ell)}=0$,
the evolution equations for ${}^2\!R$, $q_{ab}$, $\sigma^{(h)}_{ab}$
and $\theta^{(h)}$ close the evolution system
\begin{eqnarray}
\label{e:evolution_system_horizon_I}
\delta_\ell {}^2\!R &=& - \theta^{(\ell)}\,{}^2\!R 
+ 2 \; {}^2\!D^a{}^2\!D^b \sigma^{(\ell)}_{ab}
- {}^2\!\Delta \theta^{(\ell)}\,,  \\
\label{e:evolution_system_horizon_II}
\delta_\ell q_{ab} &=& 2 \sigma^{(\ell)}_{ab} +  \theta^{(\ell)} q_{ab}\,, \\
\label{e:evolution_system_horizon_III}
\delta_\ell \theta^{(\ell)} &=& -\frac{1}{2} (\theta^{(\ell)})^2
-  \sigma^{(\ell)}_{ab} {\sigma^{(\ell)}}^{ab} - 8\pi T_{ab}\ell^a\ell^b\,, \\
\label{e:evolution_system_horizon_IV}
\delta_\ell \sigma^{(\ell)}_{ab} &=&  
\sigma^{(\ell)}_{cd} {\sigma^{(\ell)}}^{cd} q_{ab} 
-  {q^c}_a{q^d}_bC_{lcfd}\ell^l\ell^f  \,.
\end{eqnarray}
Once initial conditions are prescribed, the only remaining information
needed to close the system are the matter term $T_{ab}\ell^a\ell^b$ in
the focusing equation and ${q^c}_a{q^d}_bC_{lcfd}\ell^l\ell^f$ in the
tidal equation. Using a null tetrad $(\ell^a, k^a, m^a,
\overline{m}^a)$ (see Appendix~\ref{a:geometry_Hor}) they can be
expressed in terms of Ricci and Weyl scalars: $8\pi
T_{ab}\ell^a\ell^b= R_{ab}\ell^a\ell^b= 2\Phi_{00}$ and
${q^c}_a{q^d}_bC_{lcfd}\ell^l\ell^f = \Psi_0 \overline{m}_a
\overline{m}_b + \overline{\Psi}_0 m_a m_b$. The complex Weyl scalar
$\Psi_0$ and the Ricci scalar $\Phi_{00}$ drive the evolution of the
geometric system
(\ref{e:evolution_system_horizon_I})--(\ref{e:evolution_system_horizon_IV})
on the horizon. Being determined in terms of the bulk dynamics
($\Psi_0$ relates to the near-horizon dynamical tidal fields and
incoming gravitational radiation, whereas $\Phi_{00}$ accounts for the
matter fields), fields $\Psi_0$ and $\Phi_{00}$ act as {\em external
  forces} providing (modulo initial conditions) all the relevant
dynamical information for system
(\ref{e:evolution_system_horizon_I})--(\ref{e:evolution_system_horizon_IV})
on $\scre$.

In the DH case, although the evolution system is more complex, the
qualitative conclusions reached here remain unchanged. More
specifically, the differential system on ${\cal H}$ governing the
evolution of ${}^2\!R$ is also driven by external forces given by a
particular combination of Weyl and Ricci scalars\footnote{In a DH, the
  leading term in the external driving force is indeed given by
  $\Psi_0$, but corrections proportional to $C$ also appear.}.

In the present cross-correlation approach, these dynamical
considerations strongly support $\Psi_0$ as a natural building block
in the construction\footnote{Constructed as in
  Eqs. (\ref{e:3D_Keff_v}) and (\ref{e:news_R_v}) but substituting
  ${}^2\!R$ by $\Psi_0$.}  of the quantity $h_\mathrm{inn}(v)$ at
${\cal H}$, to be correlated in vacuum to $dP_i^{\mathrm{B}}/du$ at
$\scri^+$. This is hardly surprising, given the dual nature of
$\Psi_0$ and $\Psi_4$ on inner and outer boundaries,
respectively. 

Particularly relevant are the following remarks. First, in the
presence of matter, the scalar $\Phi_{00}$ plays a role formally
analogous to that of $\Psi_0$. Therefore, in the general case, it
makes sense to consider $\Phi_{00}$ on an equal footing as $\Psi_0$ in
the construction of $h_\mathrm{inn}(v)$. Second,
Eq.~\eqref{e:evolution_system_horizon_I} is completely driven by the
rest of the system, without back-reacting on it. For this reason,
although $\Psi_0$ (and $\Phi_{00}$) encodes the information
determining the dynamics on the horizon, at the same time the
evolution of ${}^2\!R$ is sensitive to all relevant dynamical degrees
of freedom, providing an {\em averaged} response. This justifies the
crucial role of ${}^2\!R$ in the construction of the effective
$\tilde{K}_i^{\mathrm{eff}}$ in paper I.

A serious drawback for the use of $\Psi_0$ and $\Phi_{00}$ in the
construction of a quantity $h_\mathrm{inn}(v)$ at ${\cal H}$ is their
dependence on the rescaling freedom of the null normal $\ell^a$ by an
arbitrary function on ${\cal S}$. We address this point in the
following section.

\section{Fundamental results on Dynamical Horizons}
\label{s:fundamental_DH}

The introduction of a preferred null tetrad on the horizon requires
some kind of rigid structure. We argue here that DHs provide such a
structure. We first review two fundamental geometric results about
DHs:

\smallskip
a) {\em Result 1 (DH foliation uniqueness)}~\cite{Ashtekar05}: Given a
DH ${\cal H}$, the foliation ${\{\cal S}_t\}$ by marginally trapped
surfaces is unique.

\smallskip
b) {\em Result 2 (DH existence)}~\cite{ams05,AndMarSim07}: Given a
{\em strictly stably outermost} marginally trapped surface ${\cal S}_0$ in a Cauchy
hypersurface $\Sigma_0$, for each $3+1$ spacetime foliation
$\{\Sigma_t\}$ containing $\Sigma_0$ there exists a unique DH ${\cal
  H}$ containing ${\cal S}_0$ and sliced by marginally trapped
surfaces $\{ {\cal S}_t \}$ such that ${\cal S}_t\subset\Sigma_t$.

\medskip
These results have the following important implications:

\smallskip
i) {\em The evolution vector $h^a$ is completely fixed on a DH} (up to
time reparametrization). By Result 1 any other evolution vector
$h'^a$ does not transport marginally trapped surfaces into marginally
trapped surfaces.

\smallskip
ii) {\em The evolution of an AH into a DH is nonunique}. Let us
consider an initial AH ${\cal S}_0\subset \Sigma_0$ and two different
$3+1$ slicings $\{\Sigma_{t_1}\}$ and $\{\Sigma_{t_2}\}$, compatible
with $ \Sigma_0$. From Result 2 there exist DHs ${\cal H}_1 =
\bigcup_{t_1} {\cal S}_{t_1}$ and ${\cal H}_2 =\bigcup_{t_2} {\cal
  S}_{t_2} $, with ${\cal S}_{t_1} = {\cal H}_1 \cap \Sigma_{t_1}$ and
${\cal S}_{t_2} = {\cal H}_2 \cap \Sigma_{t_2}$ marginally trapped
surfaces. Let us consider now the sections of ${\cal H}_1$ by
$\{\Sigma_{t_2}\}$, \ie ${\cal S}'_{t_2} = {\cal H}_1 \cap
\Sigma_{t_2}$, so that ${\cal H}_1 = \bigcup_{t_2} {\cal
  S}'_{t_2}$. In the generic case, slicings $\{{\cal S}'_{t_2}\}$ and
$\{ {\cal S}_{t_1} \}$ of ${\cal H}_1$ are different (deform
$\{\Sigma_{t_2}\}$ if needed). Therefore, from Result 1, ${\cal
  S}'_{t_2}$ cannot be marginally trapped surfaces. Reasoning by
contradiction, we then conclude that ${\cal H}_1$ and ${\cal H}_2$ are
different hypersurfaces in ${\cal M}$, as illustrated in
Fig.\ref{fig:horizonuniq}.

\begin{figure}[t!]
\begin{center}
\includegraphics[angle=-90,width=9.0cm,]{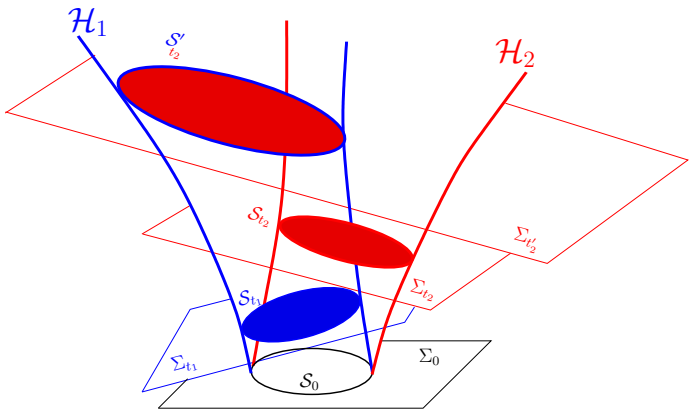}
\end{center}
\caption{ Worldtubes ${\cal H}_1$ (blue) and ${\cal H}_2$ (red),
  respectively, associated with two different $3+1$ slicing
  $\{\Sigma_{t_1}\}$ and $\{\Sigma_{t_2}\}$ and providing evolutions from
   a given marginally trapped surface ${\cal S}_0$ in an initial Cauchy
  hypersurface $\Sigma_0$. They illustrate the nonunique
  evolution of AHs into DHs. The foliation $\{ {\cal S}_{t_{1}}\}$ (resp. $\{ {\cal S}_{t_{2}}\}$ ) by
  marginally trapped surfaces is defined by the intersections of ${\cal
    H}_{1}$ with $\{\Sigma_{t_1}\}$ (resp. ${\cal
    H}_{2}$ and $\{\Sigma_{t_2}\}$). Note that, from the DH foliation uniqueness Result $1$~\cite{Ashtekar05},
  surfaces ${\cal S}'_{t_2} = {\cal H}_1 \cap \Sigma_{t_2}$ are not (in general)
  marginally trapped surfaces. } \vglue-0.5cm
\label{fig:horizonuniq}
\end{figure}

\medskip
The two results above establish a fundamental link between DHs and the
$3+1$ approach here adopted. We denote (\cf also Appendix
\ref{a:geometry_Hor}) the unit timelike normal to slices $\Sigma_t$ by
$n^a$ and the spacelike (outgoing) normal to ${\cal S}_t$ and tangent
to $\Sigma_t$ by $s^a$ (see Fig.\ref{fig:vectors}). We denote by $N$
the lapse associated to the spacetime slicing function $t$, \ie $n_a =
- N\nabla_a t$. Given a marginal trapped surface ${\cal S}_0$ in an
initial slice $\Sigma_0$, and given a lapse function $N$, let us
consider the (only) DH ${\cal H}$ given by Result 2. Then the unique
evolution vector $h^a$ on ${\cal H}$ associated with Result 1 can be
written up to a time-dependent rescaling\footnote{This applies,
  strictly, to the external part of the horizon discussed in
  Sec.~\ref{s:inner_screen}. For the internal part one must reverse
  the evolution with respect to that defined by the $3+1$ foliation:
  $h^a = -N n^a + b s^a$. The following discussion goes then through.}
as
\begin{eqnarray}
\label{e:h_n_s}
h^a = N n^a + b s^a \ \,,
\end{eqnarray} 
where $b$ is a function on ${\cal S}_t$ to be determined in terms of
$N$ and $C$ [see Eq.~(\ref{e:condition_C}) below].  Certainly such a
decomposition of an evolution vector compatible with a given $3+1$
slicing $\{\Sigma_t\}$, in the sense $h^a\nabla_at=1$, is valid for
any hypersurface but, in the case of a DH and due to Result 1, the
evolution vector $h^a$ determined by Eq.~(\ref{e:h_n_s}) has an
intrinsic meaning (up to time reparametrization, which is irrelevant
in a cross-correlation approach) as an object on ${\cal H}$ not
requiring a $3+1$ foliation. On the other hand, Eq.~(\ref{e:vector_h})
provides the expression of vector $h^a$ in terms of the null
normals. More specifically, Eq.~(\ref{e:vector_h}) links the scaling
of $\ell^a$ and $k^a$ to that of $h^a$ by imposing $h^a\to\ell^a$ as
the DH is driven to stationarity ($C\to 0\Leftrightarrow \delta_\ell
\theta^{(\ell)}\to 0$). Writing the null normals at ${\cal H}$ as
$\ell^a=(f/2)(n^a + s^a)$ and $k^a=(n^a-s^a)/f$, for some function
$f$, expressions (\ref{e:vector_h}) and (\ref{e:h_n_s}) for $h^a$ lead
to
\begin{eqnarray}
\label{e:null_normalization}
\ell_N^a = \frac{N+b}{2}\left(n^a + s^a\right) \,, \ k_N^a = \frac{1}{N+b}\left(n^a - s^a\right) \,,
\end{eqnarray}
where the subindex $N$ denotes the explicit link of ${\cal H}$ to a
$3+1$ slicing. In order to determine $b$, we evaluate the norm of
$h^a$ and note that the function $C$ in Eq.~(\ref{e:vector_h}) is
expressed in terms of $N$ and $b$ as
\begin{eqnarray}
\label{e:C_N_b}
C=\frac{1}{2}\left(b^2 - N^2\right) \ \,.
\end{eqnarray} 
On the other hand, for a given lapse $N$, the trapping horizon
$\delta_h \theta^{(\ell)}=0$ condition translates into an elliptic
equation for $C$ [\cf Equation~(\ref{e:TH_condition})]
\begin{eqnarray}
\label{e:condition_C}
&&- {}^2\!\Delta C + 2 \Omega^{(\ell)}_c  {}^2\!D^cC 
-C\left[-{}^2\!D^c  \Omega^{(\ell)}_c 
+ \Omega^{(\ell)}_c  {\Omega^{(\ell)}}^c -\frac{1}{2}{}^2\!R\right] \nn \\
&&=  \sigma^{(\ell)}_{ab} {\sigma^{(\ell)}}^{ab} + 8\pi T_{ab}\tau^a\ell^b \,. 
\end{eqnarray}
Therefore, for a given DH ${\cal H}$ associated with a $3+1$ slicing
with lapse $N$, Eqs.~(\ref{e:condition_C}) and (\ref{e:C_N_b}) fix
the value of $b$. Prescription (\ref{e:null_normalization})
provides then preferred null normals  on a DH ${\cal H}$
compatible  with the foliation defined by $N$.
Completed with the complex null vector $m^a$ on ${\cal S}_t$, we propose
\begin{eqnarray}
\label{e:canonical_null_tetrad}
(\ell^a_N, k^a_N, m^a, \overline{m}^a) \ ,
\end{eqnarray}
as a preferred null tetrad (up to time reparametrization) on a DH. To
keep the notation compact, hereafter we will denote the preferred
$\ell^a_N$ and $k^a_N$ simply as $\ell^a$ and $k^a$ and omit the
symbol $N$ from all quantities evaluated in this tetrad. The
tetrad~\eqref{e:canonical_null_tetrad} then leads to a notion of
preferred Weyl (and Ricci) scalars on the horizon ${\cal H}$. In
particular,
\begin{eqnarray}
\label{e:canonical_Psi_0_phi_0}
\Psi_0 &=&  
C^a_{\ \, bcd}\; \ell_a m^b \ell^c m^d\,, \\ 
\Phi_{00} &=& \frac{1}{2}R_{ab}\; \ell^a \ell^b \,.
\end{eqnarray}
In summary: we have introduced preferred null normals on a DH ${\cal
  H}$ by: i) linking the normalization of $\ell^a$ to that of $h^a$ by
requiring $h^a\to\ell^a$ in stationarity; and ii) fixing the
normalization of $h^a$ (up to a time-dependent function) by the
foliation uniqueness result on DHs (Result 1).  The latter is the {\em
  rigid structure} needed to fix a preferred null tetrad on ${\cal
  H}$.  In the particular case of constructing ${\cal H}$ in an
initial value problem approach (Result 2 on DHs), the free
time-dependent function is fixed by the lapse $N$ of the given global
foliation $\{\Sigma_t\}$.

\section{News-like functions 
and Bondi-like fluxes on a dynamical horizon}
\label{s:Bondi_DH_linearmomentum}

\subsection{News-like functions: vacuum case}

In Sec.~\ref{s:evol_system} we have identified the Weyl scalar
$\Psi_0$ as the object that encodes (in vacuum and for $C=0$) the
relevant geometric information on the BH horizon understood as an
inner screen. Then in Sec.  \ref{s:fundamental_DH} we have introduced
a preferred scaling for $\Psi_0$ on DHs. With these elements we can
now introduce the following vectorial quantity on ${\cal H}$
\begin{eqnarray}
\label{e:KN}
\tilde{K}[\xi](v) \equiv  - \frac{1}{8 \pi}
\oint_{{\cal S}_v} (\xi^is_i) \left|\tilde{\cal N}_{\Psi}^{(\ell)}(v)\right|^2 dA \,,
\end{eqnarray}
with
\begin{eqnarray}
\label{e:news_N_v}
\tilde{\cal N}_{\Psi}^{(\ell)}(v) 
\equiv \int_{v_0}^v \Psi_0(v') dv' \,,
\end{eqnarray}
where we make use of an advanced time $v$ parametrizing ${\cal H}$
(\cf Sec.~\ref{s:inner_screen} and Fig.~\ref{fig:advanced_time}) and
adapted to the $3+1$ slicing at ${\cal H}$ (namely, we choose
$h^a\nabla_a v=2$ to match the general notation in paper I).

The quantity $\tilde{K}_i$ could be used as a refined version of
$\tilde{K}^{\mathrm{eff}}_i$ for the correlation with
$dP_i^\mathrm{B}/du$ at $\scri^+$. However, whereas
$\tilde{K}^{\mathrm{eff}}_i$ is explicitly understood as an {\em
  effective} quantity and, consequently, one can relax the
requirement on the $\tilde{\cal N}$ constructed out of ${}^2\!R$ in
(\ref{e:news_R_v}) to behave mathematically as
a news function, the situation is different for $\tilde{K}_i$ in
(\ref{e:KN}): the geometric dual nature of $\Psi_4$ and $\Psi_0$ would
call for a newslike function character for $\tilde{\cal
  N}^{(\ell)}_{\Psi}$ in (\ref{e:news_N_v}).

Whereas expressions for the flux of Bondi momentum and the news
function at $\scri^+$ [\cf Equations~(33) and (34) in paper I] are valid
under the (strong) conditions enforced by asymptotic simplicity at
null infinity and in a given Bondi frame, no geometric structure
supports the ``a priori'' introduction of quantities $\tilde{K}_i$ and
$\tilde{\cal N}^{(\ell)}_{\Psi}$ on ${\cal H}$.  In particular, the
news function ${\cal N}(u)$ is an object well-defined in terms of
geometric quantities on sections ${\cal S}_u\subset \scri^+$, that can
be expressed as a time integral [\cf Eq.~(34) in paper I] due to the
key relation $\partial_u{{\cal N}}=\Psi_4$ holding for Bondi
coordinate systems at $\scri^+$.  On the contrary, the quantity
$\tilde{\cal N}^{(\ell)}_{\Psi}$ defined by time integration of
$\Psi_0$ is not an object defined in terms of the geometry of a
section ${\cal S}_v$ (justifying the use of a ``tilde''). Such a
local-in-time behavior is a crucial property to be satisfied by any
valid news function.  Therefore, one would expect additional terms to
$\Psi_0$ (with vanishing counterparts at $\scri^+$), contributing in
$\tilde{\cal N}^{(\ell)}_{\Psi}$ to build an appropriate newslike
function on ${\cal H}$.

In the absence of a sound geometric news formalism on ${\cal H}$, we
proceed heuristically by modifying $\tilde{\cal N}^{(\ell)}_\Psi$ so
that it acquires a local-in-time character.  Such a property would be
guaranteed if the integrand in definition (\ref{e:news_N_v}) could be
expressed as a total derivative in time of some quantity defined on
sections ${\cal S}_v$. The scalar $\Psi_0$ in Eq.~(\ref{e:news_N_v})
does not satisfy this property. However, with this guideline,
inspection of (\ref{e:evolution_system_horizon_IV}) suggests some of
the terms to be added to $\Psi_0$ [system
  (\ref{e:evolution_system_horizon_I})--(\ref{e:evolution_system_horizon_IV})
  applies to the EH case] so that they integrate in time to a quantity
on ${\cal S}_v$, namely the shear. Considering first, as an
intermediate step, the EH case and using a tensorial rather a complex
notation\footnote{We write complex numbers as $2\times 2$ traceless symmetric
  matrices.}, let us introduce a newslike tensor\footnote{Note that
  we remove now the ``tilded'' notation to emphasize its newslike
  local-in-time character.} $({\cal N}^{(\ell)}_{\Psi})_{ab}$ whose
time variation is
\be
\label{e:news_N_v_gen}
(\dot{\cal N}^{(\ell)}_{\Psi})_{ab}= \frac{1}{\sqrt{2}}
\left({q^c}_a{q^d}_bC_{lcfd}\ell^l\ell^f - 
\sigma^{(\ell)}_{cd} {\sigma^{(\ell)}}^{cd} q_{ab} \right) \,,
\ee
that is, such that $(\dot{\cal N}^{(\ell)}_{\Psi})_{ab} = -
1/\sqrt{2}\delta_{\ell} \sigma^{(\ell)}_{ab}$ (the global factor
$1/\sqrt{2}$ is required for the correct coefficient in the
leading-order contribution).  Upon time integration in Eq. (\ref{e:news_N_v})
and setting vanishing initial values at early times, this choice leads
to
\begin{eqnarray}
\label{e:news_event_horizon}
({\cal N}^{(\ell)}_{\Psi})_{ab} = -
\frac{1}{\sqrt{2}}\sigma^{(\ell)}_{ab} \ \ \,. \ \ \
\end{eqnarray}
If we write
\bea
\label{e:int_N}
({\cal N}^{(\ell)}_{\Psi})_{ab} &=& \frac{1}{\sqrt{2}} \int_{v_0}^v 
\left[{q^c}_a{q^d}_bC_{lcfd}\ell^l\ell^f - \right. \nn \\
&&\left. 2 ({\cal N}^{(\ell)}_{\Psi})_{cd}({{\cal N}^{(\ell)}}_{\Psi})^{cd} q_{ab} \right]dv' \,,
\eea
and substitute $({\cal N}^{(\ell)}_{\Psi})_{ab}$ recursively in the
right hand side, we can express the newslike function $({\cal
  N}^{(\ell)}_{\Psi})_{ab}$ in terms of $\Psi_0$ so that the
lowest-order term is indeed given by expression (\ref{e:news_N_v}).

This identification, in the EH case, of a plausible newslike tensor
as the shear along the evolution vector suggests the following
specific proposal for the newslike tensor for DHs
\begin{eqnarray}
\label{e:news_DH_grav}
{\cal N}^{^{({\cal H})}}_{ab} \equiv - \frac{1}{\sqrt{2}} \sigma^{(h)}_{ab} \,.
\end{eqnarray}
This proposal has a tentative character. Once we have identified the basics,
we postpone a systematic study to a forthcoming work.

\subsection{News-like functions: matter fields}

As discussed in Sec.~\ref{s:evol_system}, in system
(\ref{e:evolution_system_horizon_I})--(\ref{e:evolution_system_horizon_IV})
the Ricci scalar $\Phi_{00}$ plays a role analogous to that of
$\Psi_0$. From this perspective, in the matter case, it is reasonable
to define as in (\ref{e:news_N_v})
\begin{eqnarray}
\label{e:News_H_matter}
\tilde{\cal N}^{(\ell)}_{\Phi}(v) \equiv \frac{\alpha_{\mathrm{m}}}{2}
 \int_{v_0}^v \Phi_{00}(v') \; dv' \,,
\end{eqnarray}
such that $\tilde{K}_i$ in (\ref{e:KN}) is rewritten
\be
\label{e:KN_matter}
\tilde{K}[\xi](v) \equiv  - \frac{1}{8 \pi}
\oint_{{\cal S}_v} (\xi^is_i) 
\left[\left|\tilde{\cal N}_{\Psi}^{(\ell)}(v)\right|^2 
+ \left(\tilde{\cal N}_{\Phi}^{(\ell)}(v) \right)^2\right] dA \,.
\ee
The parameter $\alpha_{\mathrm{m}}$ is introduced to account for
possible different relative contributions of $\Psi_0$ and $\Phi_{00}$
(distinct choices for $\alpha_{\mathrm{m}}$ are possible, depending on
the particular quantity to be correlated at $\scri^+$).  However, also
the function $\tilde{\cal N}^{(\ell)}_{\Phi}$ is affected by the same
issues discussed above for $\tilde{\cal N}^{(\ell)}_{\Psi}$, namely it
lacks a local-in-time behavior.  As in the vacuum case, we proceed
first by looking at EHs.  We then complete $\Phi_{00}$ with the terms
in Eq.~(\ref{e:evolution_system_horizon_III}), so that $\dot{\cal
  N}^{(\ell)}_{\Phi}(v) = - ({\alpha_{\mathrm{m}}}/{2}) \delta_\ell
\theta^{(\ell)}$. That is
\begin{eqnarray}
\label{e:new_matter}
\dot{\cal N}^{(\ell)}_{\Phi}(v)=
\frac{\alpha_{\mathrm{m}}}{2}
\left( 8\pi T_{cd}\ell^c\ell^d +\frac{1}{2} (\theta^{(\ell)})^2
+ \sigma^{(\ell)}_{cd} {\sigma^{(\ell)}}^{cd}  \right) \nn \,,
\end{eqnarray}
so that ${\cal N}^{(\ell)}_{\Phi}=- (\alpha_{\mathrm{m}}/2)
\theta^{(\ell)}$.  This {\em matter newslike} function can be
equivalently expressed in tensorial form as follows
\begin{eqnarray}
\label{e:new_matter_tensor}
({\cal N}^{(\ell)}_{\Phi})_{ab}= -\frac{\alpha_{\mathrm{m}}}{2\sqrt{2}}
\theta^{(\ell)} q_{ab} \ .
\end{eqnarray}

As in vacuum, the passage from EHs to DHs is accomplished by using the
natural evolution vector $h^a$ along ${\cal H}$ for the
expansion. Then, combining the tensorial form
(\ref{e:new_matter_tensor}) with (\ref{e:news_DH_grav}), we can write
a single newslike tensor as
\begin{eqnarray}
\label{e:news_N}
{\cal N}^{^{({\cal H})}}_{ab}= - \frac{1}{\sqrt{2}}
\left(
\sigma^{(h)}_{ab}
+\frac{\alpha_{\mathrm{m}}}{2} \theta^{(h)} q_{ab} \right) \ .
\end{eqnarray}
Interestingly, if $\alpha_{\mathrm{m}}=1$ the complete news tensor
acquires a clear geometric meaning as the deformation tensor along
$h^a$, \ie as the time variation of the induced metric
\begin{eqnarray}
\label{e:news_N_Theta}
{\cal N}^{^{({\cal H})}}_{ab}= - \frac{1}{\sqrt{2}} \Theta^{(h)}_{ab}=
 - \frac{1}{2\sqrt{2}} \dot{q}_{ab} \,.
\end{eqnarray}

\subsection{Bondi-like fluxes on ${\cal H}$}

The motivation for introducing $\tilde{K}_i^{\mathrm{eff}}$ in paper I
and $\tilde{K}_i$ in Eq.~(\ref{e:KN}) [or, more generally,
  $\tilde{K}_{i}$ in Eq.~(\ref{e:KN_matter})] is the construction of
quantities on ${\cal H}$ to be correlated to quantities at $\scri^+$,
namely the flux of Bondi linear momentum.  We have been careful not to
refer to them as to ``fluxes,'' since they do not have an
instantaneous meaning.  However, once the newslike tensor ${\cal
  N}^{^{({\cal H})}}_{ab}$ has been introduced in (\ref{e:news_N}),
formal fluxes can be constructed by integration of the squared of
these news. More specifically, we can introduce the formal fluxes on
${\cal H}$
\begin{eqnarray}
\label{e:fluxes_DH}
\frac{dE^{^{({\cal H})}}}{dv}(v) &=&
\frac{1}{8 \pi}
\oint_{{\cal S}_v} {\cal N}^{^{({\cal H})}}_{ab}{\cal N}^{^{({\cal H})} ab}dA \,,  \\
\frac{dP^{^{({\cal H})}}[\xi]}{dv}(v) &=&
-\frac{1}{8 \pi}
\oint_{{\cal S}_v} (\xi^i s_i)
\left({\cal N}^{^{({\cal H})}}_{ab}{\cal N}^{^{({\cal H})} ab}\right) \nn
dA  \,, 
\end{eqnarray}
where their formal notation as total time derivatives is meant to make
explicit their local-in-time nature. The purpose of quantities
$dE^{{^{({\cal H})}}}/dv$ and $(dP^{{^{({\cal H})}}}[\xi]/dv)$ is to provide improved quantities
at ${\cal H}$ for the cross-correlation approach. In particular,
$(dP^{{^{({\cal H})}}}[\xi]/dv)$ provides a refined version of the effective
$\tilde{K}_i^{\mathrm{eff}}$ in paper I, to be correlated with
$(dP_i^{\mathrm{B}}/du)(u)$ at $\scri^+$.  In this context,
$\tilde{K}_i$ in Eq.~(\ref{e:KN}) has played the role of an
intermediate stage in our line of arguments.

Of course, we can introduce formal quantities $E^{{^{({\cal H})}}}$ and
$P^{{^{({\cal H})}}}_i$ on ${\cal H}$, by integrating expressions in
(\ref{e:fluxes_DH}) along ${\cal H}$.  However, in the absence of a
physical conservation argument or a geometric motivation, referring to
them as (Bondi-like) energies and momentum would be just a matter of
definition\footnote{For instance, the leading-order contribution from
  matter to the BH energy and momentum should come from the
  integration of the appropriate component of the stress-energy tensor
  $T_{ab}$, an element absent in (\ref{e:fluxes_DH}) where matter
  contributions only enter through terms quadratic in
  $T_{ab}$.}. Thus, we rather interpret them simply as well-defined
instantaneous quantities leading ultimately to a timeseries
$h_{\mathrm{inn}}(v)$.

It is illustrative to expand the squared of the news in
(\ref{e:fluxes_DH}) as
\begin{eqnarray}
\label{e:squared_news}
{\cal N}^{^{({\cal H})}}_{ab}{\cal N}^{^{({\cal H})} ab}
=\frac{1}{2}\left[\sigma^{(h)}_{ab}{\sigma^{(h)}}^{ab}
+ \frac{\alpha_{\mathrm{m}}}{2} (\theta^{(h)})^2 \right] ,
\end{eqnarray}
to be inserted in the expression for $dE^{{^{({\cal H})}}}/dv$ and
$(dP^{{^{({\cal H})}}}[\xi]/dv)$.  The relative weight of the different terms as we
depart from equilibrium can be made explicit by expressing the
evolution vector as $h^a= \ell^a-Ck^a$ [\cf Equation~(\ref{e:vector_h})],
with associated $\sigma^{(h)}_{ab} =
\sigma^{(\ell)}_{ab}-C\sigma^{(k)}_{ab}$ and $\theta^{(h)} =
-C\theta^{(k)}$ [\cf Equation~\eqref{e:theta_sigma_h}]. We can then write
\begin{eqnarray}
\label{e:squared_news_C}
{\cal N}^{^{({\cal H})}}_{ab}{\cal N}^{^{({\cal H})} ab}
&=&\frac{1}{2}\left[\sigma^{(\ell)}_{ab}{\sigma^{(\ell)}}^{ab}
 -2C\sigma^{(\ell)}_{ab}{\sigma^{(k)}}^{ab} \right. \nn\\
&&\left. +C^2\left(\sigma^{(k)}_{ab}{\sigma^{(k)}}^{ab}
+ \frac{\alpha_{\mathrm{m}}}{2}(\theta^{(k)})^2\right)\right].
\end{eqnarray}
On a DH, terms proportional to $\alpha_{\mathrm{m}}$ only enter at a
quadratic order in $C$.  Two values of $\alpha_{\mathrm{m}}$ are of
particular interest.  First, the case $\alpha_{\mathrm{m}}=0$,
corresponding to an analysis of pure gravitational dynamics. Second,
the case $\alpha_{\mathrm{m}}=1$ where [\cf (\ref{e:news_N_Theta})]
\begin{eqnarray}
\label{e:kinetic}
\frac{dE^{{^{({\cal H})}}}}{dv}(v)&=&
\frac{1}{8 \pi}
\oint_{{\cal S}_v} {\cal N}^{^{({\cal H})}}_{ab}{\cal N}^{^{({\cal H})} ab}dA  = 
\frac{1}{16 \pi}\oint_{{\cal S}_v} \Theta^{(h)}_{ab}{\Theta^{(h)}}^{ab}dA  \nn\\
&=&
\frac{1}{32 \pi} \oint_{{\cal S}_v}\frac{1}{2}(\dot{q}_{ab})^2dA  \,,
\end{eqnarray}
that admits a suggestive interpretation as a  Newtonian
{\em kinetic energy} term of the intrinsic horizon geometry.

\subsection{Relation to quasilocal approaches to horizon momentum
and application to recoil dynamics} 
\label{s:quasilocal_momentum}

As emphasized in the previous section, the essential purpose of
$dE^{{^{({\cal H})}}}/dv$ and $(dP^{{^{({\cal H})}}}[\xi]/dv)$ in (\ref{e:fluxes_DH}) is to
provide geometrically sound proposals for $h_{\mathrm{inn}}(v)$ at
${\cal H}$.  Having said this, it is worthwhile to compare the
resulting expressions, for specific values of $\alpha_{\mathrm{m}}$,
with DH physical fluxes derived in the literature.  This provides an
internal consistency test of the line of thought followed from
$\tilde{K}_i^{\mathrm{eff}}$ to Eqs. (\ref{e:fluxes_DH}). In
particular, for $\alpha_{\mathrm{m}}=0$ we obtain
\begin{widetext}
\be
\label{e:E_shear_expansion}
\frac{dE^{{^{({\cal H})}}}}{dv}(v) =
\frac{1}{8 \pi}
\oint_{{\cal S}_v} {\cal N}^{^{({\cal H})}}_{ab}{\cal N}^{^{({\cal H})}  ab} dA   
=
\frac{1}{16 \pi}
\oint_{{\cal S}_v}  \sigma^{(h)}_{ab}{\sigma^{(h)}}^{ab}
dA   
 = \frac{1}{16 \pi}
\oint_{{\cal S}_v}  \left[\sigma^{(\ell)}_{ab}{\sigma^{(\ell)}}^{ab}
 -2C\sigma^{(\ell)}_{ab}{\sigma^{(k)}}^{ab} 
+C^2 \sigma^{(k)}_{ab}{\sigma^{(k)}}^{ab}\right]dA  \ \,.
\ee
\end{widetext}

Expression~\eqref{e:E_shear_expansion} allows us to draw analogies
with the energy flux proposed in the DH geometric analysis of
Refs.~\cite{Ashtekar-etal-2002-dynamical-horizons, Ashtekar03a}.  In
particular, the leading term in the integrand of this expression,
$\sigma^{(\ell)}_{ab}{\sigma^{(\ell)}}^{ab}$, is directly linked [\cf
Equation~(3.27) in~\cite{Ashtekar03a}] to the term identified
in~\cite{Hay04,Hayward04} as the flux of {\em transverse}
gravitational propagating degrees of freedom\footnote{We note that
  $\oint_{\cal S} \sigma^{(\ell)}_{ab}{\sigma^{(\ell)}}^{ab}dA$ was
  used in Ref.~\cite{JarVasAns07} as a practical dimensionless
  parameter to monitor horizons approaching stationarity. Here they
  would correspond to a vanishing flow of transverse radiation.}.  The
DH energy flux also includes a {\em longitudinal} part~\cite{Hay04,Hayward04} 
depending on $\Omega^{(\ell)}_a{\Omega^{(\ell)}}^a$, absent in quantities in
Eq.~(\ref{e:fluxes_DH}). In this sense, $dE^{N}/dv$ provides a
quantity $h_{\mathrm{inn}}(v)$ accounting only for the transverse part
of gravitational degrees of freedom~\cite{Szekeres65, Nolan04,
  Hayward04} at ${\cal H}$ and therefore particularly suited for
cross-correlation with $(dP_i^{\mathrm{B}}/du)(u)$, which corresponds
to (purely transverse) gravitational radiation at $\scri^+$.

Motivated now by the resemblance of (\ref{e:E_shear_expansion}) with
the flux of a physical quantity, we can consider a heuristic notion of
{\em Bondi-like} 4-momentum flux through ${\cal H}$.  Considering the
(timelike) unit normal $\hat{\tau}^a$ to ${\cal H}$ [\cf (\ref{e:h_m})
  and (\ref{e:h_N_b})]
\be
\label{e:tauhat}
\hat{\tau}^a= \frac{\tau^a}{\sqrt{|\tau^b\tau_b|}}=
\frac{1}{\sqrt{2C}} (\ell^a+Ck^a)= \frac{1}{\sqrt{2C}} (bn^a + Ns^a) \ ,
\ee
we can introduce the component of a 4-momentum flux $(dP_a^{\tau}/dv)$
along a generic 4-vector $\eta^a$, as
\begin{eqnarray}
\label{e:Bondi_4momentum}
\frac{dP^\tau[\eta]}{dv} &\equiv& -\frac{1}{8 \pi}
\oint_{{\cal S}_v} (\eta^c \hat{\tau}_c) 
\left({\cal N}^{^{({\cal H})}}_{ab}{\cal N}^{^{({\cal H})} ab}\right)dA \nn \\
&= &-\frac{1}{16 \pi}
\oint_{{\cal S}_v} (\eta^c \hat{\tau}_c) \left({\sigma}^{(h)}_{ab}
{\sigma^{(h)}}^{ab}\right)dA  \ ,
\end{eqnarray}
that has formally the expression of the flux of a {\em Bondi-like} 4-momentum.
The corresponding flux of {\em energy} associated with an Eulerian 
observer $n^a$ is
\begin{eqnarray}
\label{e:energy_tau}
\frac{dE^{\tau}}{dv}(v) \equiv \frac{dP^\tau[n]}{dv}  =
\frac{1}{16 \pi}
\oint_{\cal S} \frac{b}{\sqrt{2C}}\left(\sigma^{(h)}_{ab}{\sigma^{(h)}}^{ab}\right)
dA  \ \,, \nn \\
\end{eqnarray}
where $\frac{b}{\sqrt{2C}}=\sqrt{1+N^2/2C}$. Analogously, the flux of
linear momentum for $\xi^a$ tangent to $\Sigma_t$ would be 
\be
\label{e:momentum_tau}
\frac{dP^\tau[\xi]}{dv} = -\frac{1}{16 \pi}
\oint_{{\cal S}_v} \frac{N}{\sqrt{2C}}(\xi^i s_i) 
\left(\sigma^{(h)}_{ab}{\sigma^{(h)}}^{ab}\right)dA.
\ee
Near equilibrium, \ie for $C\to0$, we have
${\sigma}^{(h)}_{ab}{{\sigma}^{(h)}}^{ab}\sim C$ on DHs [\cf
  Equations~(\ref{e:C_spherical})] so that the integrands in expressions
(\ref{e:energy_tau}) and (\ref{e:momentum_tau}) are $O(\sqrt{C})$,
therefore regular and vanishing in this limit. Considering
$(dP^\tau[\xi]/dv)$ as an estimate of the flux of gravitational linear
momentum\footnote{A related alternative prescription for a DH linear
  momentum flux would be given by angular integration of the
  appropriate components in the {effective gravitational-radiation
    energy-tensor} in~\cite{Hayward04}.} through ${\cal H}$, the
integrated quantity $P_i^\tau$ would provide a {\em heuristic} prescription for a
quasilocal DH linear momentum, a sort of {\em Bondi}-like counterpart
of the heuristic {\em ADM}-like linear momentum introduced for DHs in
Ref.~\cite{Krishnan:2007pu}, by applying the ADM expression for the
linear momentum at spatial infinity $i^o$ to the DH section ${\cal
  S}_t$
\begin{eqnarray}
\label{e:ADM-like_momentum}
P[\xi] = \frac{1}{8\pi}\int_{{\cal S}_t} \left(K_{ab} -K \gamma_{ab}\right)
\xi^a s^b \; dA \ \, .
\end{eqnarray}
In this sense, the {cross-correlation} methodology we propose here and
in paper I, can be formally compared with the quasilocal momentum
approaches in Refs.~\cite{Krishnan:2007pu,Lovelace:2009dg} to the
study of the recoil velocity in binary BHs mergers, showing the
complementarity among these lines of research.
 
However, attempting to derive in our context a rigorous notion of
quasilocal momentum on ${\cal H}$ would require the development of a
systematic news-functions {\em framework} on DHs, in particular
considering the possibility of longitudinal gravitational terms as in
the DH energy flux (\cf Refs.~\cite{Wu06,  Wu:2009dm,
  Wu:2010eu} for important insights in this topic).  Such a discussion
is beyond our present heuristic treatment, and we stick to our
approach of considering the constructed local fluxes on ${\cal H}$ as
quantities encoding information about (transverse) propagating
gravitational degrees of freedom to be cross-correlated to the flux of
Bondi momentum at $\scri^+$.

\section{Link to the Horizon viscous-fluid picture}
\label{s:viscous_picture}

The basic idea proposed in Ref.~\cite{Rezzolla:2010df} is that certain
qualitative aspects of the late-time BH recoil dynamics, and, in
particular, the antikick, can be understood in terms of the
dissipation of the anisotropic distribution of curvature on the
horizon. This picture in which the BH recoils as a result of the
emission of anisotropic gravitational radiation in response to an
anisotropic curvature distribution suggests that the interaction of
the moving BH with its environment induces a {\em viscous} dissipation
of the gravitational dynamics. 
The cross-correlation approach to near-horizon dynamics 
discussed  in paper I and complemented here 
offers a realization of the idea proposed
in~\cite{Rezzolla:2010df}, expressing it in more geometrical terms.
Indeed, the analysis in 
 Sec.~\ref{s:Bondi_DH_linearmomentum} has led us to the identification of the
shear $ \sigma^{(h)}_{ab}$ and of the expansion $\theta^{(h)}$, 
interpreted there in terms of newslike functions at ${\cal H}$,
as the relevant objects in tracking the geometry evolution.
This identification permits to cast naturally the viscous-fluid picture into a more sound
basis, since $\theta^{(h)}$ and
$\sigma^{(h)}_{ab}$ have indeed an interpretation in terms of bulk
and shear viscosities. 
Such dissipative features can already be appreciated
explicitly in Eq.~(\ref{e:R_evolution}), but acquire a larger basis in
the context of the membrane paradigm that we review below.

\subsection{The BH horizon viscous-fluid analogy}
\label{s:membrane_paradigm}

Hawking and Hartle~\cite{Hawking:1972hy,Hartle:1973zz,Hartle:1974gy}
introduced the notion of {\em BH viscosity} when studying the response
of the event horizon to external perturbations. This leads to a
viscous-fluid analogy for the treatment of the physics of the EH,
fully developed by Damour~\cite{Damour79,Damour82} and by Thorne,
Price and Macdonald~\cite{Price86, Price86a}, in the so-called {\em
  membrane paradigm} (see also \cite{Straumann:1997tg,
  Damour:2008ji}). In this approach, the physical properties of the BH
are discussed in terms of mechanical and electromagnetic properties of
a 2-dimensional viscous fluid. A quasilocal version of some of its
aspects, applying for dynamical trapping horizons, has been developed
in \cite{Gourgoulhon-Jaramillo-Review, Gourgoulhon:2005ch,
  Gourgoulhon:2006uc, Gourgoulhon:2008pu}.

In the fluid analogy of the membrane paradigm, dissipation in BH
dynamics is accounted for in terms of the shear and bulk viscosities
of the fluid. The viscosity coefficients are identified in the
dissipative terms appearing in the momentum and energy balance
equations for the 2-dimensional fluid. These equations are obtained
from the projection of the appropriate components of the Einstein
equations on the horizon's world-tube, namely evolution equations for
$\theta^{(\ell)}$ and $\Omega^{(\ell)}_a$. For an EH these equations
are~\cite{Gourgoulhon-Jaramillo-Review}
\begin{eqnarray}
\label{e:theta_Omega_EH}
\delta_\ell \theta^{(\ell)} -  \kappa^{(\ell)}\theta^{(\ell)}&=&
-{1\over 2} {\theta^{(\ell)}}^2 
- \sigma^{(\ell)}_{ab} {\sigma^{(\ell)}}^{ab} 
	- 8\pi T_{ab}\ell^a\ell^b \ , \nn \\
\delta_\ell \Omega^{(\ell)}_a + \theta^{(\ell)} \, \Omega^{(\ell)}_a &=&       
       {}^2\!D_a \left( \kappa^{(\ell)} + \frac{\theta^{(\ell)}}{2} \right)
       -  {}^2\!D_c {\sigma^{(\ell)}}^c_{\ \, a}  \nn \\
       &&+8\pi T_{cd} \, \ell^c q^d_{\ \, a} \, .
\end{eqnarray}
The first one [\ie the Raychaudhuri
Eq.~(\ref{e:evolution_system_horizon_III}] not assuming a affine
geodesic parametrization, so that $\kappa^{(\ell)}\neq 0$) is
interpreted as an energy dissipation equation. In particular, a {\em
  surface energy density} is identified as $\varepsilon \equiv
-{\theta^{(\ell)}}/{8\pi}$. The second evolution equation for the
normal form $\Omega^{(\ell)}_a$ provides a momentum conservation
equation for the fluid, a Navier-Stokes-like equation (referred to as
Damour-Navier-Stokes equation), once a momentum $\pi_a$ for the
2-dimensional fluid is identified as $\pi_a \equiv -\Omega^{(\ell)}_a
/ (8\pi)$ [note that $\Omega^{(\ell)}_a$ is associated with a density
  of angular momentum; \cf Equation~(\ref{e:angular_momentum})]. Dividing
Eqs. (\ref{e:theta_Omega_EH}) by ${-8\pi}$ and applying these
identifications we obtain
\begin{widetext}
\begin{eqnarray}
\label{e:energy_DNS_equations_EH}
\delta_\ell \varepsilon + \theta^{(\ell)} \varepsilon &=&
-\left(\frac{\kappa^{(\ell)}}{8\pi}\right)  \theta^{(\ell)}  - \frac{1}{16\pi} (\theta^{(\ell)})^2
+ \sigma^{(\ell)}_{cd} \left(\frac{{\sigma^{(\ell)}}^{cd}}{8\pi}\right) + 
T_{ab}\ell^a\ell^b \ ,\\ 
\delta_\ell \pi_a + \theta^{(\ell)}
\pi_a & =& -{}^2\!D_a\left(\frac{\kappa^{(\ell)}}{8\pi}\right) 
+ {}^2\!D^c\left(\frac{\sigma^{(\ell)}_{ca}}{8\pi}\right) 
-{}^2\!D_a\left(\frac{\theta^{(\ell)}}{16\pi}\right) 
- {q^c}_aT_{cd}\ell^d \ .
\end{eqnarray} 
\end{widetext}
Writing the null evolution vector as $\ell^a=\partial_t +V^a$, for some
(velocity) vector $V^a$ tangent to ${\cal S}_t$, one can write $
\theta^{(\ell)} = D_aV^a+\partial_t \mathrm{ln}\sqrt{q}$ and
$2\sigma^{(\ell)}_{ab}=\left( {}^2\!D_aV_b+{}^2\!D_bV_a\right)
-\theta^{(\ell)}q_{ab} + \frac{1}{2}\partial_tq_{ab}$. Then one can
identify a fluid pressure $P\equiv {\kappa^{(\ell)}}/{(8\pi)}$, a (negative) bulk
viscosity coefficient $\zeta = -{1}/{(16\pi)}$, a shear viscosity
coefficient $\mu = 1/(16\pi)$, an external energy production rate
$T_{ab}\ell^a\ell^b $ and external force density $f_a\equiv-
{q^c}_aT_{cd}\ell^d$ . See also~\cite{Padma11} for a criticism of this
interpretation.

The analogue equations in dynamical trapping horizons are obtained
from the equations $\delta_h \theta^{h}$ and $\delta_h\Omega^{(\ell)}_a$.
The latter can be written as~\cite{Gourgoulhon:2005ch, Gourgoulhon:2006uc,
  Gourgoulhon:2008pu}
\begin{widetext}
\begin{eqnarray}
\label{e:delta_h_theta_h}
\left(\delta_{h} + \theta^{(h)}\right)\theta^{(h)}  &=&
	- \kappa^{(h)} \theta^{(h)} 
	+  \sigma^{(h)}_{ab} {\sigma^{(\tau)}}^{ab}
	+ \frac{(\theta^{(h)})^2}{2}   
	+ {}^2\!D^a ( {}^2\!D_aC- 2C \Omega^{(\ell)}_a ) 
	+ 8\pi T_{ab}\tau^a h^b  
	- \theta^{(k)} \delta_hC  \ \ , \\
\label{e:delta_h_Omega}
\left(\delta_{h} + \theta^{(h)}\right) \Omega^{(\ell)}_a 
	& =&  {}^2\!D_a \kappa^{(h)}
	- {}^2\!D^c \sigma^{(\tau)}_{ac}
	- \frac{1}{2}{}^2\!D_a\theta^{(h)} 
         + 8\pi {q^b}_aT_{bc}\tau^c 
	- \theta^{(k)} {}^2\!D_a C \ .
\end{eqnarray}
\end{widetext}
with $\kappa^{(h)}= -k_ah^b \nabla_b \ell^a$ [see
  Eq.~(\ref{e:kappa_v})]. Then, by introducing a {\em DH surface
  energy density} $\bar{\varepsilon} \equiv
-{\theta^{(\tau)}}/{(8\pi)}= {\theta^{(h)}}/{(8\pi)}$, keeping the
definition for $\pi_a$ and introducing the {\em heat} 
$Q_a = \frac{1}{4\pi} \left[ C \Omega^{(\ell)}_a - \frac{1}{2}
  {}^2\!D_aC \right]$, we can write for DHs
(see~\cite{Gourgoulhon:2006uc, Gourgoulhon:2008pu} for a complete
interpretation of these equations)
\begin{widetext}
\begin{eqnarray}
\label{e:energy_DNS_equations_DH}
\delta_{h}\bar{\varepsilon} + \theta^{(h)} \bar{\varepsilon}  &=& 
	- \left(\frac{\kappa^{(h)}}{8\pi}\right) \theta^{(h)} 
	+ \frac{1}{16\pi}(\theta^{(h)})^2
	+   \sigma^{(h)}_{ab} \left( \frac{{\sigma^{(\tau)}}^{ab}}{8\pi}\right)
	+ T_{ab}\tau^a h^b  
        -  {}^2\!D^a Q_a 
	- \frac{\theta^{(k)}}{8\pi} \delta_hC  \ \,, \nn \\
\delta_{h} \pi_a 
	+ \theta^{(h)} \pi_a & = & -{}^2\!D_a \left(\frac{\kappa^{(h)}}{8\pi} 
\right)
	+ {}^2\!D^c \left( 
	\frac{\sigma^{(\tau)}_{ac}}{8\pi} \right)
	+ {}^2\!D_a\left(\frac{\theta^{(h)}}{16\pi}\right)  
        - {q^b}_aT_{bc}\tau^c 
	+ \frac{\theta^{(k)}}{8\pi} {}^2\!D_a C \ .
\end{eqnarray}
\end{widetext}
We can now justify the viscosity interpretation of $\theta^{(h)}$ and
$\sigma^{(h)}_{ab}$ by remarking that from the equations above,
$\theta^{(h)}$ represents the expansion of the fluid in the
bulk viscosity term [with positive bulk viscosity coefficient
  $\zeta=1/{(16\pi)}$]. Similarly, $\sigma^{(h)}_{ab}$ corresponds to
the shear strain tensor and $\sigma^{(\tau)}_{ab}/(8\pi)$ to the shear
stress tensor. Note that $\sigma^{(\tau)}_{ab}/(8\pi)$ and
$\sigma^{(h)}_{ab}$ are not proportional in the strict dynamical case,
$C\neq 0$, and therefore one cannot define a shear viscosity
coefficient $\mu$ (in other words, a DH is not a Newtonian fluid).

Finally let us consider the observer given by the (properly
normalized) timelike normal to ${\cal H}$ and let us define the
4-momentum current density associated with this observer:
$p_a\equiv-T_{ab}\tau^b$. Then we note that the components of $p_a$
are fixed by Eqs. (\ref{e:energy_DNS_equations_DH}) together with the
trapping horizon defining constraint Eq.~(\ref{e:condition_C}).
Indeed, $p_a h^a = - T_{ab}\tau^b h^a$ corresponds to the energy
$\bar{\varepsilon}$ balance equation, while $p_b {q^b}_a =
-T_{bc}\tau^c{q^b}_a$ gives the momentum $\pi_a$ conservation equation,
and $p_a \tau^a = - T_{ab}\tau^b \tau^a$ is a linear combination,
using $\tau^a= 2\ell^a - h^a$, of the energy dissipation equation and
 the trapping horizon condition ($\delta_h
\theta^{(\ell)}=0$) depending on $T_{ab} \tau^a\ell^b$. 
Given the fundamental role of the latter in the
geometric properties of the DH, in particular in the derivation of an
area law under the future condition $\theta^{(k)}<0$, this suggests
the possibility of using the component $p_a \tau^a$ to define a
balance equation for an appropriate entropy density. This point echoes
the discussion of a hydrodynamic entropy current discussed in the context of a
fluid-gravity duality~\cite{BhaMinHub08, BhaHubLog08, Booth:2010kr,
  BooHelPle11,Eling:2009sj}.

\subsection{A viscous ``slowness parameter''}

The viscosity interpretation outlined in the previous subsection
allows us now to make contact with the {\em slowness parameter} $P$
introduced in~\cite{Price:2011fm} and discussed in paper I in the
context of BH head-on collisions. We recall that the parameter $P$ is
constructed in terms of two dynamical timescales: a decay timescale
$\tau$ and an oscillation time scale $T$
\begin{eqnarray}
\label{e:slowness_param}
P=\frac{T}{\tau}.
\end{eqnarray}
In our fluid analogy, the bulk viscosity term $\theta^{(h)}$ controls
the dynamical decay, whereas the shear viscosity term
$\sigma^{(h)}_{ab}$ is responsible for the (shape) oscillations of the
geometry. Given their physical dimensions $[\theta^{(h)}] =
[\sigma^{(h)}_{ab}] = [\mathrm{Length}]^{-1}$, averaging over horizon
sections we can build instantaneous timescales\footnote{These are not
  the only possibility to define $\tau$ and $T$, and therefore $P$,
  from viscosity scales. All variants should give though the same
  qualitative estimates; see Eq.~(\ref{e:viscosity_timescales2}).} at
any coordinate time $t$ as
\begin{eqnarray}
\label{e:viscosity_timescales_I}
\frac{1}{\tau(t)} &\equiv& 
\frac{1}{A}\oint_{{\cal S}_t}(\xi_t^is_i)\theta^{(h)}dA \ ,  \\
\label{e:viscosity_timescales_II}
\frac{1}{T(t)^2} &\equiv&\frac{1}{A}\oint_{{\cal S}_t} 
(\xi_t^is_i)\left({\sigma}^{(h)}_{ab}{\sigma^{(h)}}^{ab}\right)dA \ ,
\end{eqnarray}
where $\xi_t^i$ is the unit vector in the instantaneous direction of
motion of the BH at time $t$. The term $(\xi_t^is_i)$ in the
definitions~\eqref{e:viscosity_timescales_I} -
\eqref{e:viscosity_timescales_II} is needed for giving a timescale
associated with a change in {linear momentum} [if not, we would be
  dealing with a timescale for a change in {energy}, \cf
  (\ref{e:fluxes_DH})]. In other words, it is needed to account for
the dissipation and oscillation of anisotropies in the geometry 
rather than for spherically symmetric growths. This is consistent with the
beating-frequency behavior found in the timeseries developed for the
head-on collision of two BHs (\cf Eq.~(58) in paper I). Note that
Eqs.~(\ref{e:viscosity_timescales_I})-(\ref{e:viscosity_timescales_II})
provide geometric prescriptions for the instantaneous timescales at
the merger of a binary system, an open problem pointed out in~\cite{Price:2011fm}.
Combining Eqs.~(\ref{e:slowness_param})
and~(\ref{e:viscosity_timescales_I})-(\ref{e:viscosity_timescales_II}),
and denoting $|{\sigma}^{(h)}|^2 = {\sigma}^{(h)}_{ab}
{\sigma^{(h)}}^{ab}$, we get
\begin{eqnarray}
\label{e:P_theta_sigma}
P(t)=\frac{\oint_{{\cal S}_t}(\xi_t^is_i)\theta^{(h)}dA }
{\left[A \, \oint_{{\cal S}_t} 
(\xi_t^is_i)|{\sigma}^{(h)}|^2dA\right]^{\frac{1}{2}}} \ .
\end{eqnarray}
As a consistency check we can verify for DHs that using
Eqs.~(\ref{e:theta_sigma_h}) and (\ref{e:C_spherical}) and in
situations close to stationarity (\ie $C\to 0$), the following scaling
holds $\theta^{(h)}\sim C$ and $|{\sigma}^{(h)}|^2\sim C$, so that
$P$ remains well-defined in this limit. For an alternative and
more sound proposal for $P$, improving further the behavior
when $C\to 0$, see Eq.~(\ref{e:viscosity_timescales2}).

\section{Conclusions}
\label{s:conclusions}

The analysis of spacetime dynamics is a very hard task in the absence
of some rigid structure, such as symmetries or a preferred background
geometry. However, this is the generic situation in the strong-field
regime described by general relativity. In this context,
(complementary) {\em effective} approaches providing insight into the
qualitative aspects of the solutions and suggesting avenues for their
quantitative modeling are of much value. In this spirit, in paper I
and here, we have discussed a cross-correlation approach to
near-horizon dynamics. Other interesting schemes, such as those
developed at Caltech, that define and exploit new
curvature-visualization tools~\cite{OweBriChe11,Nichols2011}, share
some aspects of this methodological approach.

In particular, we have argued that, in the setting of a $3+1$ approach
to the BH spacetime construction, the foliation uniqueness of
dynamical horizons provides a rigid structure that confers a preferred
character to these hypersurfaces as probes of the BH
geometry. Employed as inner screens in the {\em cross-correlation}
approach, this DH foliation uniqueness permits to introduce the
preferred normalization (\ref{e:null_normalization}) of the null
normals to AH sections and, consequently, a preferred angular scaling
in the Weyl scalars on these horizons. The remaining time
reparametrization freedom ({time-stretch issue}) does not affect the
adopted cross-correlation scheme, where only the structure of the
respective sequence maxima and minima is of relevance in the
correlation of quantities defined at outer and inner screens.

Although this natural scaling of the Weyl tensors on DHs has an
interest of its own, we have employed it here as an intermediate
stage, linking the effective-curvature vector
$\tilde{K}^{\mathrm{eff}}_i(t)$ in paper I to the identification of
the shear $\sigma^{(h)}_{ab}$, associated with the DH evolution vector
$h^a$, as being proportional to a geometric DH newslike function
${\cal N}^{^{({\cal H})}}_{ab}$ in Eq.~(\ref{e:news_DH_grav}) [see
  also the role of $\theta^{(h)}$, in the more general ${\cal
    N}^{^{({\cal H})}}_{ab}$ in Eq.~(\ref{e:news_N})]. On the one
hand, this identification provides a (refined) geometric flux quantity
$(dP^{{^{({\cal H})}}}_i/dv)$ on DH sections to be correlated to
the flux of Bondi linear momentum $(dP^{\mathrm{B}}_i/du)$ at
$\scri^+$ (these DH fluxes also share features with quasilocal linear
momentum treatments in the literature). On the other hand, given the
role of $\sigma^{(h)}_{ab}$ and $\theta^{(h)}$ in driving the Ricci
scalar ${}^2\!R$ along ${\cal H}$ [namely Eq. (\ref{e:R_evolution})
  and system
  (\ref{e:evolution_system_horizon_I})-(\ref{e:evolution_system_horizon_IV})],
the present analysis justifies the use of
$\tilde{K}^{\mathrm{eff}}_i(t)$ in paper I as an effective local
estimator at ${\cal H}$ of dynamical aspects at $\scri^+$.

The cross-correlation analysis has also produced two important
by-products. First, we advocate the physical relevance of tracking the
internal horizon in $3+1$ BH evolutions. This follows from the
consideration of the time integration of fluxes along the horizon and
its splitting (\ref{e:News_t_splitting}) into internal horizon and
external horizon integrals (cf. Appendix~\ref{appendixB}). Such expression is fixed up to an
early-times integration constant, controlled by dynamics previous to
the formation of the (common) DH (and possibly vanishing in many
situations of interest). Second and most importantly, from the
perspective of a viscous-horizon analogy we have identified a
dynamical decay timescale $\tau$ associated with bulk viscosity and an
oscillation timescale $T$ associated with the shear viscosity [\cf
  Equations.~\eqref{e:viscosity_timescales_I}-\eqref{e:viscosity_timescales_II}
  and also Eqs.~(\ref{e:viscosity_timescales2})]. This is particularly
relevant in the context of BH recoil dynamics, where the analysis
in~\cite{Price:2011fm} shows that the qualitative features of the late-time
recoil can be explained in terms of a generic behavior
controlled by the relative values of a decay and an oscillation
time scales. The viscous picture meets the rationale
in~\cite{Rezzolla:2010df} and offers an understanding of the relevant
dynamical time scales from the (trace and traceless parts in the)
evolution of the horizon intrinsic geometry, in particular, providing
{\em instantaneous} dynamical time scales at the merger and a geometric
prescription [\cf Equation~(\ref{e:P_theta_sigma}) and also
  Eq.~(\ref{e:P_theta_sigma_2})] for the slowness parameter $P=T/\tau$
introduced in~\cite{Price:2011fm}.

As a final remark we note that while the material presented here
places the arguments made in~\cite{Rezzolla:2010df} and in paper I on
a much more robust geometrical basis, much of our treatment is still
heuristic and based on intuition. More work is needed for the
development of a fully systematic framework and this will be the
subject of our future research.


\begin{acknowledgments}
It is a pleasure to thank S. Bai, S. Bonazzola, E. Gourgoulhon,
C.-M. Cheng, B. Krishnan, J. Novak, J. Nester, A. Nielsen, A. Tonita,
C.-H. Wang, Y.-H. Wu, B. Schutz and J.M.M. Senovilla for useful
discussions. This work was supported in part by the DAAD and the DFG
grant SFB/Transregio~7. J.L.J. acknowledges support from the Alexander
von Humboldt Foundation, the Spanish MICINN (FIS2008-06078-C03-01) and
the Junta de Andaluc\'\i a (FQM2288/219).
\end{acknowledgments} 

\appendix

\section{A geometric brief }
\label{a:geometry_Hor}

We bring together in this Appendix the different geometric objects and
structures that have been introduced in the text, on the spacetime
$({\cal M}, g_{ab})$ with Levi-Civita connection $\nabla_a$.

\subsection{Geometry of sections ${\cal S}_t$}

\noi {\em Normal plane to ${\cal S}$}. Given a spacelike closed
(compact without boundary) 2-surface ${\cal S}$ in ${\cal M}$ and a
point $p\in {\cal S}$, the normal plane to ${\cal S}$, $T_p^\perp
{\cal S}$, can be spanned by (future-oriented) null vectors $\ell^a$
and $k^a$ (defined by the intersection between $T_p^\perp {\cal S}$
and the null cone at $p$). We choose a normalization $\ell^a k_a =
-1$. Directions $\ell^a$ and $k^a$ are uniquely determined, but a {\em
  normalization-boost} freedom remains: $\ell'^a = f \ell^a$, $k'^a =
f^{-1} k^a$.

\noi {\em Intrinsic geometry on ${\cal S}$.}
The induced metric on ${\cal S}$ is given by
\be
\label{e:metric_S}
q_{ab}=g_{ab}+k_a \ell_b + \ell_a k_b \ .
\ee
We denote the Levi-Civita connection associated with $q_{ab}$ as
${}^2\!D_a$. The area form on ${\cal S}$ is given
by ${}^2\!\epsilon = \sqrt{q} dx^1\wedge dx^2$, \ie
${}^2\!\epsilon_{ab}= k^c\ell^d{}^4\!\epsilon_{cdab}$, and 
we use the area measure notation $dA=\sqrt{q}d^2x$.

\noi {\em Extrinsic geometry of ${\cal S}$.}
First, given a vector $v^a$ orthogonal to ${\cal S}$, 
we denote the derivative at ${\cal S}$
of a tensor ${X^{a_1\ldots a_n}}_{b_1\ldots b_m}$ tangent
to ${\cal S}$ along $v^a$, as 
\begin{eqnarray}
\label{e:delta_v_X}
&&\delta_v {X^{a_1\ldots a_n}}_{b_1\ldots b_m} \equiv \\
&&{q^{a_1}}_{c_1}\ldots{q^{a_n}}_{c_n} 
{q^{d_1}}_{b_1}\ldots{q^{d_m}}_{b_m} 
{\cal L}_v {X^{c_1\ldots c_n}}_{d_1\ldots d_m} \ , \nn
\end{eqnarray}
where 
${\cal L}_v$ denote the Lie derivative along (some extension of) $v^a$.
Then, the {\em deformation tensor} $\Theta^{(v)}_{ab}$
along a vector $v^a$ normal to ${\cal S}$
\be
\label{e:deformation_tensor}
\Theta^{(v)}_{ab} \equiv  {q^c}_a {q^d}_b \nabla_c v_d 
= \frac{1}{2} {\delta}_v q_{ab} \ , 
\ee
encodes the deformation of the intrinsic geometry along $v^a$.
More generally, the {\em second fundamental tensor} is defined as
\be
\label{e:second_fund_form_S}
{\cal K}^c_{ab}\equiv {q^d}_a {q^e}_b \nabla_d {q^c}_e = k^c \Theta^{(\ell)}_{ab} + \ell^c \Theta^{(k)}_{ab} \ \,.
\ee
We can express $\Theta^{(v)}_{ab}$ in terms of its trace and traceless parts 
\begin{eqnarray}
\label{e:trace_traceless_Theta}
\Theta^{(v)}_{ab} = \sigma^{(v)}_{ab} +  \frac{1}{2} \theta^{(v)} q_{ab} \ \,,
\end{eqnarray}
where $\theta^{(v)}$ and  $\sigma^{(v)}_{ab}$ denote, respectively,
the expansion and shear along $v^a$
\be
\label{e:expansion_shear}
\theta^{(v)} \equiv q^{ab}\nabla_av_b =\frac{1}{\sqrt{q}}\delta_v \sqrt{q}
\ \,, \ \ 
\sigma^{(v)}_{ab} \equiv \Theta^{(v)}_{ab} - \frac{1}{2} \theta^{(v)} q_{ab} \ \,.
\ee
Information on the extrinsic geometry of  $({\cal S}, q_{ab})$ in 
$({\cal M}, g_{ab})$ is completed by the  {\em normal form} 
$\Omega^{(\ell)}_a$, defined as
\be
\label{e:normal_form}
\Omega^{(\ell)}_a \equiv -k^c {q^d}_a \nabla_d \ell_c \ \,.
\ee
In particular,  given an axial Killing vector $\phi^a$ on ${\cal S}$, an
angular momentum $J[\phi]$ (coinciding with the Komar angular momentum
if $\phi^a$ can be extended to a Killing vector in the neighborhood of
${\cal S}$) can be defined as
\begin{eqnarray}  
\label{e:angular_momentum}
J[\phi]=\frac{1}{8\pi}\int_{\cal S} \Omega^{(\ell)}_a\phi^a dA \ .
\end{eqnarray}
This quantity is well-defined for any divergence-free axial vector $\phi^a$.
Finally, given a vector $v^a\in T^\perp {\cal S}$ we define
\cite{Gourgoulhon:2005ch} 
\begin{eqnarray}
\label{e:kappa_v}
\kappa^{(v)} = - k_a v^c\nabla_c \ell^a \ .
\end{eqnarray}

\noi {\em Remark on $\delta_v \theta^{(\ell)}$}. In
(\ref{e:delta_v_X}) we have introduced $\delta_v$ in terms of the Lie
derivative on tensorial objects. However, the evaluations of
expressions such as $\delta_v \theta^{(\ell)}$ is more delicate, since
$\theta^{(\ell)}$ is not a scalar quantity on ${\cal M}$, but rather a
quasilocal object depending on ${\cal S}$. In the general case,
$\delta_{\gamma v}\theta^{(\ell)}$ (with $\gamma$ a function on ${\cal
  S}$) depends on the deformation induced on ${\cal S}$ by $\gamma$,
so that $\delta_{\gamma v}\theta^{(\ell)} \neq
\gamma\delta_{v}\theta^{(\ell)}$. This is the reason for the special
notation $\delta_v$. Properties $\delta_{a v + b w}\theta^{(\ell)}=
a\delta_v \theta^{(\ell)}+ b \delta_w\theta^{(\ell)}$ ($a, b \in
\mathbb{R}$), and the Leibnitz rule $\delta_v (\gamma \theta^{(\ell)})
= (\delta_v \gamma) \theta^{(\ell)} + \gamma \delta_v \theta^{(\ell)}$
still hold. See for instance
Refs.~\cite{ams05,Booth:2006bn,Cao:2010vj} for a discussion of this
derivative operator.

\subsection{Evolution on the horizon ${\cal H}$}

Given a DH ${\cal H}$, it has a unique foliation $\{{\cal S}_t\}$ by
marginally trapped surfaces. This fixes, up to {\em time}
reparametrization, the evolution vector $h^a$ along ${\cal H}$. This is
characterized as being tangent to ${\cal H}$ and orthogonal to ${\cal
  S}_t$, and Lie-transporting ${\cal S}_t$ onto ${\cal S}_{t+\delta
  t}$: ${\delta}_h t =1$. We write $h^a$ and a {\em dual} vector
$\tau^a$ orthogonal to ${\cal H}$ in terms of the null normals as
\begin{eqnarray}
\label{e:h_m}
h^a = \ell^a - C k^a \ \ , \ \ \tau^a = \ell^a + C k^a \ .
\end{eqnarray}
Then $h^a h_a = -\tau^a \tau_a = 2C$.
The expansion $\theta^{(h)}$ and shear $\sigma^{(h)}_{ab}$ are written as 
\begin{eqnarray}
\label{e:theta_sigma_h}
\theta^{(h)}&=& \theta^{(\ell)} - C \theta^{(k)} = - C \theta^{(k)} \ , \nn \\
\sigma^{(h)}_{ab}&=& \sigma^{(\ell)}_{ab}-C\sigma^{(k)}_{ab} \ .
\end{eqnarray}
The DH is characterized by $\theta^{(\ell)}=0$
and  $\delta_h \theta^{(\ell)}=0$. Using (\ref{e:h_m})
and the properties of the $\delta_v$ operator,
the latter condition is expressed as 
\begin{eqnarray}
\label{e:TH_condition}
- {}^2\!\Delta C + 2 \Omega^{(\ell)}_c  {}^2\!D^cC -
C\delta_{k} \theta^{(\ell)}= -\delta_\ell \theta^{(\ell)}  \ , 
\end{eqnarray}
an elliptic equation on $C$. Under the {\em outer} condition
in~\ref{s:BHs}, $\delta_k \theta^{(\ell)}<0$, a maximum principle can
be applied so that $C\geq 0$, with $C=0$ if and only if $\delta_\ell
\theta^{(\ell)}=0$ (stationary case). Therefore, a (future outer)
trapping horizon ${\cal H}$ is fully partitioned in purely stationary
and purely dynamical sections. In other words, sections of ${\cal H}$
react {\em as a whole}, growing in size everywhere as soon as some
energy crosses the horizon somewhere. This nonlocal elliptic
behavior is inherited from the defining trapping horizon condition
Eq.~(\ref{e:TH_condition}). Substituting
\begin{eqnarray}
\label{e:Aell_Bk}
\delta_{\ell} \theta^{(\ell)} &=& -\left(\sigma^{(\ell)}_{ab} {\sigma^{(\ell)}}^{ab}
 - 8\pi T_{ab}\ell^a\ell^b\right) \ , \\
\delta_{k} \theta^{(\ell)} &=& 
-{}^2\!D^c  \Omega^{(\ell)}_c 
+ \Omega^{(\ell)}_c  {\Omega^{(\ell)}}^c -\frac{1}{2}{}^2\!R
+ 8\pi T_{ab}k^a\ell^b 
 \nn \ ,
\end{eqnarray}
into (\ref{e:TH_condition}), we recover Eq.~(\ref{e:condition_C}) in the text. 
In the spherically symmetric case ($C=\mathrm{const}$), and
using the expression for $\delta_{\ell} \theta^{(\ell)}$
in (\ref{e:Aell_Bk}) into (\ref{e:TH_condition}),
we get
\be
\label{e:C_spherical}
C = - \frac{\sigma_{ab}^{(\ell)}{\sigma^{(\ell)}}^{ab}+ 8\pi T_{ab}\ell^a\ell^b}
{\delta_k \theta^{(\ell)} } \ \ .
\ee

\subsection{3+1 perspective on the horizon ${\cal H}$}

Given a $3+1$ foliation of spacetime $\{ \Sigma_t \}$ defined by a {\em
  time} function $t$, we denote the unit timelike normal to $\Sigma_t$
by $n^a$ and the lapse function by $N$, \ie $n_a=-N\nabla_at$. The
induced metric on $\Sigma_t$ is denoted by $\gamma_{ab}$, \ie
$\gamma_{ab} = g_{ab} + n_a n_b $ with Levi-Civita connection
$D_a$. The extrinsic curvature of $\Sigma_t$ in ${\cal M}$ is $K_{ab}
= -{\gamma^c}_a\nabla_c n_b$. We consider a horizon ${\cal H}$, such
that the spacetime foliation $\{ \Sigma_t \}$ induces a foliation
$\{{\cal S}_t\}$ of ${\cal H}$ by marginal trapped surfaces. From
Result 1 in Sec.~\ref{s:fundamental_DH} this foliation is unique.
Let us denote the normal to ${\cal S}_t$ tangent to $\Sigma_t$ by
$s^a$. Vectors $n^a$ and $s^a$ span also the normal plane to ${\cal
  S}_t$. From the condition ${\delta}_h t =1$ we can write $h^a$ and
$\tau^a$ in (\ref{e:h_m}) as
\begin{eqnarray}
\label{e:h_N_b}
h^a = N n^a + b s^a \ \ , \ \  \tau^a = b n^a + N s^a \ ,
\end{eqnarray}
for some function $b$ on ${\cal S}_t$, expressed 
in terms of $N$ and $C$ in (\ref{e:h_m}), as
$2C=(b+N)(b-N)$. 

\subsection{An improved geometric prescription for the slowness
parameter}
\label{Pdef_2}

In Eqs.~(\ref{e:viscosity_timescales_I}) -
(\ref{e:viscosity_timescales_II}) we have introduced decay and
oscillation instantaneous timescales 
from $\theta^{(h)}$ and $\sigma^{(h)}_{ab}$, respectively, identified as
newslike functions at ${\cal H}$ in Sec.~\ref{s:Bondi_DH_linearmomentum}
and responsible for bulk and
shear viscosities on ${\cal H}$ (\cf Sec. \ref{s:viscous_picture}).
This is not the only possibility. From the bulk and shear viscosity terms in
Eq.~(\ref{e:delta_h_theta_h}) we define
\begin{eqnarray}
\label{e:viscosity_timescales2}
\frac{1}{\tau(t)^2}&\equiv& \frac{1}{A}
\oint_{{\cal S}_t}(\xi_t^is_i)\left(\kappa^{(h)}\theta^{(h)}\right) dA \ ,\nn \\
\frac{1}{T(t)^2}&\equiv& \frac{1}{A}
\oint_{{\cal S}_t}(\xi_t^is_i)
\left(\sigma_{ab}^{(h)}{\sigma^{(\tau)}}^{ab}\right) dA \ ,
\end{eqnarray}
where $\kappa^{(h)}$ can be expressed, in a $3+1$ decomposition, as
\be
\label{e:kappah_3+1}
\kappa^{(h)}=Ns^aD_a N - b s^as^bK_{ab}
+\delta_h \mathrm{ln}\left(\frac{N+b}{2}\right).
\ee
Then, the slowness parameter $P=T/\tau$ in Eq.~(\ref{e:slowness_param})
results
\begin{eqnarray}
\label{e:P_theta_sigma_2}
P(t)=\left(\frac{\oint_{{\cal S}_t}(\xi_t^is_i)
\left(\kappa^{(h)}\theta^{(h)}\right) dA}
{\oint_{{\cal S}_t}(\xi_t^is_i)\left(\sigma_{ab}^{(h)}{\sigma^{(\tau)}}^{ab}\right) 
dA}\right)^{\frac{1}{2}}.
\end{eqnarray}
Note that, neglecting derivative and high-order terms in
Eq.~(\ref{e:delta_h_theta_h}) near stationarity ($C\to 0$), we get
$\kappa^{(h)}\theta^{(h)}\sim\sigma_{ab}^{(h)}{\sigma^{(\tau)}}^{ab}$,
so that $P\sim 1$ consistently with the expected absence of {\em
  antikick} in this limit (cf.~\cite{Price:2011fm}).

\subsection{Weyl and Ricci scalars}

Let us complete null vectors $\ell^a$ and $k^a$ in $T^\perp{\cal S}_t$
to a tetrad $\{ \ell^a, k^a, (e_1)^a, (e_2)^a \}$, where $(e_i)^a$ are
orthonormal vectors tangent to ${\cal S}_t$. Defining the complex
null vector $m^a = \frac{1}{\sqrt{2}}[(e_1)^a+ i (e_2)^a]$, the Weyl
scalars are defined as the components of the Weyl tensor $C^a_{\ \,
  bcd}$ in the null tetrad $\{ \ell^a, k^a, m^a, \overline{m}^a \}$
\be
\label{e:Weyl_scalars_psis}
	\begin{array}{ll}
\Psi_0 = C^a_{\ \, bcd}\; \ell_a m^b \ell^c m^d, & 
\qquad \Psi_3 = C^a_{\ \, bcd}\; \ell_a k^b \overline{m}^c k^d, \\
\Psi_1 = C^a_{\ \, bcd}\; \ell_a m^b \ell^c k^d, &
\qquad \Psi_4 = C^a_{\ \, bcd}\; \overline{m}_a k^b \overline{m}^c k^d, \\
\Psi_2 = C^a_{\ \, bcd}\; \ell_a m^b \overline{m}^c k^d. &
    	\end{array}
\ee
Ricci scalars are then defined as
\be
\label{e:Weyl_scalars_phis}
	\begin{array}{ll}
\Phi_{00} = -\frac{1}{2}R_{ab}\ell^a\ell^b, & 
\Phi_{21} = -\frac{1}{2}R_{ab}k^a\overline{m}^b,  \\
\Phi_{11} = -\frac{1}{4}R_{ab}\left(\ell^ak^b+m^a\overline{m}^b\right), & 
\Phi_{02} = -\frac{1}{2}R_{ab}m^am^b, \\
\Phi_{01} = -\frac{1}{2}R_{ab}\ell^am^b,   & 
\Phi_{22} = -\frac{1}{2}R_{ab}k^ak^b,  \\
\Phi_{12} = -\frac{1}{2}R_{ab}k^am^b, & 
\Phi_{20} = -\frac{1}{2}R_{ab}\overline{m}^a\overline{m}^b, \\
\Phi_{10} = -\frac{1}{2}R_{ab}\ell^a\overline{m}^b, & 
\Lambda = \frac{1}{24}R. \\
    	\end{array}
\ee

\section{Relevance of the 3+1 inner common horizon}
\label{appendixB}

In this appendix we emphasize the role of the inner horizon present in
$3+1$ slicings of BH spacetimes, and discussed in
Sec.~\ref{s:inner_screen}, when considering the time integration of
fluxes along the DH history. This is of specific relevance to the
discussion made in Sec.~\ref{s:quasilocal_momentum}, but it also applies to
more general contexts.

Given a flux density $\frac{dQ}{dAdv}(\Omega,v)$ through ${\cal H}$ of a physical quantity
$Q(v)$, we can write
\begin{eqnarray}
\label{e:flux_Q}
Q(v)&=&Q(v_0) +  \int_{v_0}^v \left(\oint_{{\cal S}_v} \mathrm{sign}(C) \frac{dQ}{dAdv}(\Omega, v')dA\right) 
dv' \nn \\
&=& Q(v_0) +  \int_{v_0}^v F_C(v')dv'
\ ,
\end{eqnarray}
where\footnote{The sign $\mathrm{sign}(C)$,
  $+1$ for spacelike and $-1$ for timelike sectors of ${\cal H}$,
  corrects the possibility of integrating twice (null) fluxes through
  ${\cal H}$, when timelike parts occur in the world-tube 
of the trapping horizon ${\cal H}$. Note that $\mathrm{sign}(C)$
appears under the integral since a section ${\cal S}_v$
can be partially timelike and partially spacelike,
i.e. the evolution vector $h^a$ can be timelike or spacelike in differemt
parts of ${\cal S}_v$.} $F_C(v)\equiv \oint_{{\cal S}_v} \mathrm{sign}(C) \frac{dQ}{dAdv}(v')dA$.
This requires a good parametrization of ${\cal H}$ by the
(advanced) coordinate  $v$, as well as an initial value $Q(v_0)$. Finding
such an initial value is in general nontrivial and this is precisely
the motivation to consider in this section the evaluation of the
fluxes along the whole spacetime history of ${\cal H}$, though
from a $3+1$ perspective.

Given the $3+1$ slicing $\{\Sigma_t\}$, we can split the integration
along the DH into an external and an internal horizon
parts, as discussed in Sec.~\ref{s:inner_screen}. Denoting by $v_c$
the advanced time associated with the moment $t_c$ of first $3+1$
appearance of the horizon, ${\cal H}$ is separated into the inner
horizon ${\cal H}_{\mathrm{int}}$ labeled by $v_0\leq v\leq v_c$ and
the outer horizon ${\cal H}_{\mathrm{out}}$ labeled by $v_c\leq
v\leq \infty$: ${\cal H} = {\cal H}_{\mathrm{int}} \cup {\cal
  H}_{\mathrm{ext}} = \left(\bigcup_{v_0\leq v\leq v_c}{\cal
  S}_v\right) \cup \left(\bigcup_{v_c\leq v\leq \infty}{\cal
  S}_v\right)$. We can then rewrite Eq.~(\ref{e:flux_Q}) as
\begin{widetext}
\begin{eqnarray}
\label{e:flux_Q_v_splitting}
Q(v)&=&Q(v_0) + \int_{v_0}^vF_C(v')dv'  
= Q(v_0) + \int_{v_0}^{v_c} F_C^{\mathrm{int}}(v') dv' 
+  \int_{v_c}^v F_C^{\mathrm{ext}}(v') dv' \\
\label{e:flux_Q_v_splitting_2}
&=& Q(v_0) + 
\int_{v_c}^{2v_c -v_0} F_C^{\mathrm{int}}(2v_c - v'') dv'' 
+\int_{v_c}^v  F_C^{\mathrm{ext}}(v') dv' \,,
\end{eqnarray}
\end{widetext}
where $F_C^{\mathrm{int}}$ and
$F_C^{\mathrm{ext}}$ denote, respectively, the flux of $Q$
along the internal and external horizons. Note that in the second term
in~\eqref{e:flux_Q_v_splitting_2} we have inverted the integration
limits in order to have an expression which is ready to be translated
for an integration in $t$.

The coordinate $v$ is not usually adopted in standard $3+1$
numerical constructions of spacetimes. Because of this, we employ the
time $t$ defining the slicing $\{\Sigma_t\}$. Although the $t$
function is not a good parameter on the whole ${\cal H}$, it correctly
parametrizes the evolution of both the inner ${\cal H}_{\mathrm{int}}$
and outer ${\cal H}_{\mathrm{ext}}$ horizons separately: ${\cal H} =
{\cal H}_{\mathrm{int}} \cup {\cal H}_{\mathrm{ext}} =
\left(\bigcup_{t\geq t_c}{\cal S}^{\mathrm{int}}_t\right) \cup
\left(\bigcup_{t\geq t_c}{\cal S}^{\mathrm{ext}}_t\right)$.
Considering the splitting in Eq.~(\ref{e:flux_Q_v_splitting}), the use
of $t$ in the flux integration is perfectly valid as long as the
$t$-integration includes both the standard external horizon part {\em
  and} an internal horizon part. 

From Eq. (\ref{e:flux_Q_v_splitting_2}) we write
\begin{widetext}
\begin{eqnarray}
\label{e:News_t_splitting}
Q(t)&=&Q_0 
+  \int^{\infty}_{t_c} F_C^{\mathrm{int}}(t') dt'
+  \int_{t_c}^t F_C^{\mathrm{ext}}(t') dt' \nn \\
&=&Q_0 
+  \int^{t}_{t_c}  F_C^{\mathrm{int}}(t') dt'
+  \int_{t_c}^t F_C^{\mathrm{ext}}(t') dt' +
\mathrm{Res}(t) \ , 
\end{eqnarray}
\end{widetext}
where $Q_0$ is a constant and the error $\mathrm{Res}(t)$ 
\begin{eqnarray}
\label{e:error_news_t}
\mathrm{Res}(t)= \int^{\infty}_{t}F_C^{\mathrm{int}}(t') dt' ,
\end{eqnarray}
must be taken into account, since we cannot integrate up to
$t\to\infty$ during the $3+1$ evolution. This error satisfies
$\mathrm{Res}(t)\to 0$ as $t\to \infty$, so that the evaluation of
$Q(t)$ by ignoring $\mathrm{Res}(t)$ in Eq.~(\ref{e:News_t_splitting})
improves as we advance in time $t$ (\cf Figure~\ref{fig:advanced_time}).
Of course, this approach requires a good numerical tracking of the
inner horizon, something potentially challenging from a numerical
point of view (see~\cite{Szilagyi:2006qy} for a related discussion).

\bibliographystyle{apsrev4-1-noeprint.bst}
\bibliography{aeireferences}

\begin{thebibliography}{55}%
\makeatletter
\providecommand \@ifxundefined [1]{%
 \@ifx{#1\undefined}
}%
\providecommand \@ifnum [1]{%
 \ifnum #1\expandafter \@firstoftwo
 \else \expandafter \@secondoftwo
 \fi
}%
\providecommand \@ifx [1]{%
 \ifx #1\expandafter \@firstoftwo
 \else \expandafter \@secondoftwo
 \fi
}%
\providecommand \natexlab [1]{#1}%
\providecommand \enquote  [1]{``#1''}%
\providecommand \bibnamefont  [1]{#1}%
\providecommand \bibfnamefont [1]{#1}%
\providecommand \citenamefont [1]{#1}%
\providecommand \href@noop [0]{\@secondoftwo}%
\providecommand \href [0]{\begingroup \@sanitize@url \@href}%
\providecommand \@href[1]{\@@startlink{#1}\@@href}%
\providecommand \@@href[1]{\endgroup#1\@@endlink}%
\providecommand \@sanitize@url [0]{\catcode `\\12\catcode `\$12\catcode
  `\&12\catcode `\#12\catcode `\^12\catcode `\_12\catcode `\%12\relax}%
\providecommand \@@startlink[1]{}%
\providecommand \@@endlink[0]{}%
\providecommand \url  [0]{\begingroup\@sanitize@url \@url }%
\providecommand \@url [1]{\endgroup\@href {#1}{\urlprefix }}%
\providecommand \urlprefix  [0]{URL }%
\providecommand \Eprint [0]{\href }%
\providecommand \doibase [0]{http://dx.doi.org/}%
\providecommand \selectlanguage [0]{\@gobble}%
\providecommand \bibinfo  [0]{\@secondoftwo}%
\providecommand \bibfield  [0]{\@secondoftwo}%
\providecommand \translation [1]{[#1]}%
\providecommand \BibitemOpen [0]{}%
\providecommand \bibitemStop [0]{}%
\providecommand \bibitemNoStop [0]{.\EOS\space}%
\providecommand \EOS [0]{\spacefactor3000\relax}%
\providecommand \BibitemShut  [1]{\csname bibitem#1\endcsname}%
\let\auto@bib@innerbib\@empty
\bibitem [{\citenamefont {Jaramillo}\ \emph {et~al.}(2012)\citenamefont
  {Jaramillo}, \citenamefont {Macedo}, \citenamefont {Moesta},\ and\
  \citenamefont {Rezzolla}}]{Jaramillo:2011re}%
  \BibitemOpen
  \bibfield  {author} {\bibinfo {author} {\bibfnamefont {J.~L.}\ \bibnamefont
  {Jaramillo}}, \bibinfo {author} {\bibfnamefont {R.~P.}\ \bibnamefont
  {Macedo}}, \bibinfo {author} {\bibfnamefont {P.}~\bibnamefont {Moesta}}, \
  and\ \bibinfo {author} {\bibfnamefont {L.}~\bibnamefont {Rezzolla}},\ }\href
  {\doibase 10.1103/PhysRevD.85.084030} {\bibfield  {journal} {\bibinfo
  {journal} {Phys. Rev. D}\ }\textbf {\bibinfo {volume} {85}},\ \bibinfo
  {pages} {084030} (\bibinfo {year} {2012})}\BibitemShut {NoStop}%
\bibitem [{\citenamefont {Rezzolla}\ \emph {et~al.}(2010)\citenamefont
  {Rezzolla}, \citenamefont {Macedo},\ and\ \citenamefont
  {Jaramillo}}]{Rezzolla:2010df}%
  \BibitemOpen
  \bibfield  {author} {\bibinfo {author} {\bibfnamefont {L.}~\bibnamefont
  {Rezzolla}}, \bibinfo {author} {\bibfnamefont {R.~P.}\ \bibnamefont
  {Macedo}}, \ and\ \bibinfo {author} {\bibfnamefont {J.~L.}\ \bibnamefont
  {Jaramillo}},\ }\href {\doibase 10.1103/PhysRevLett.104.221101} {\bibfield
  {journal} {\bibinfo  {journal} {Phys. Rev. Lett.}\ }\textbf {\bibinfo
  {volume} {104}},\ \bibinfo {pages} {221101} (\bibinfo {year}
  {2010})}\BibitemShut {NoStop}%
\bibitem [{\citenamefont {Hayward}(1994{\natexlab{a}})}]{Hayward94a}%
  \BibitemOpen
  \bibfield  {author} {\bibinfo {author} {\bibfnamefont {S.~A.}\ \bibnamefont
  {Hayward}},\ }\href {\doibase 10.1103/PhysRevD.49.6467} {\bibfield  {journal}
  {\bibinfo  {journal} {Phys. Rev. D}\ }\textbf {\bibinfo {volume} {49}},\
  \bibinfo {pages} {6467} (\bibinfo {year} {1994}{\natexlab{a}})}\BibitemShut
  {NoStop}%
\bibitem [{\citenamefont {Ashtekar}\ and\ \citenamefont
  {Krishnan}(2003)}]{Ashtekar03a}%
  \BibitemOpen
  \bibfield  {author} {\bibinfo {author} {\bibfnamefont {A.}~\bibnamefont
  {Ashtekar}}\ and\ \bibinfo {author} {\bibfnamefont {B.}~\bibnamefont
  {Krishnan}},\ }\href@noop {} {\bibfield  {journal} {\bibinfo  {journal}
  {Phys. Rev. D}\ }\textbf {\bibinfo {volume} {68}},\ \bibinfo {pages} {104030}
  (\bibinfo {year} {2003})}\BibitemShut {NoStop}%
\bibitem [{\citenamefont {Ashtekar}\ and\ \citenamefont
  {Krishnan}(2004)}]{Ashtekar:2004cn}%
  \BibitemOpen
  \bibfield  {author} {\bibinfo {author} {\bibfnamefont {A.}~\bibnamefont
  {Ashtekar}}\ and\ \bibinfo {author} {\bibfnamefont {B.}~\bibnamefont
  {Krishnan}},\ }\href@noop {} {\bibfield  {journal} {\bibinfo  {journal}
  {Living Rev. Relativ.}\ }\textbf {\bibinfo {volume} {7}},\ \bibinfo {pages}
  {10} (\bibinfo {year} {2004})}\BibitemShut {NoStop}%
\bibitem [{\citenamefont {Hayward}(1994{\natexlab{b}})}]{Haywa94c}%
  \BibitemOpen
  \bibfield  {author} {\bibinfo {author} {\bibfnamefont {S.~A.}\ \bibnamefont
  {Hayward}},\ }\href {http://stacks.iop.org/0264-9381/11/i=12/a=017}
  {\bibfield  {journal} {\bibinfo  {journal} {Classical Quantum Gravity}\
  }\textbf {\bibinfo {volume} {11}},\ \bibinfo {pages} {3037} (\bibinfo {year}
  {1994}{\natexlab{b}})}\BibitemShut {NoStop}%
\bibitem [{\citenamefont {Hayward}(2003)}]{Haywa03}%
  \BibitemOpen
  \bibfield  {author} {\bibinfo {author} {\bibfnamefont {S.~A.}\ \bibnamefont
  {Hayward}},\ }\href {\doibase 10.1103/PhysRevD.68.104015} {\bibfield
  {journal} {\bibinfo  {journal} {Phys. Rev. D}\ }\textbf {\bibinfo {volume}
  {68}},\ \bibinfo {pages} {104015} (\bibinfo {year} {2003})}\BibitemShut
  {NoStop}%
\bibitem [{\citenamefont {{Price}}\ \emph {et~al.}(2011)\citenamefont
  {{Price}}, \citenamefont {{Khanna}},\ and\ \citenamefont
  {{Hughes}}}]{Price:2011fm}%
  \BibitemOpen
  \bibfield  {author} {\bibinfo {author} {\bibfnamefont {R.~H.}\ \bibnamefont
  {{Price}}}, \bibinfo {author} {\bibfnamefont {G.}~\bibnamefont {{Khanna}}}, \
  and\ \bibinfo {author} {\bibfnamefont {S.~A.}\ \bibnamefont {{Hughes}}},\
  }\href {\doibase 10.1103/PhysRevD.83.124002} {\bibfield  {journal} {\bibinfo
  {journal} {Phys. Rev. D}\ }\textbf {\bibinfo {volume} {83}},\ \bibinfo
  {pages} {124002} (\bibinfo {year} {2011})}\BibitemShut {NoStop}%
\bibitem [{\citenamefont {Booth}\ and\ \citenamefont
  {Fairhurst}(2004)}]{Booth04a}%
  \BibitemOpen
  \bibfield  {author} {\bibinfo {author} {\bibfnamefont {I.}~\bibnamefont
  {Booth}}\ and\ \bibinfo {author} {\bibfnamefont {S.}~\bibnamefont
  {Fairhurst}},\ }\href@noop {} {\bibfield  {journal} {\bibinfo  {journal}
  {Phys. Rev. Lett}\ }\textbf {\bibinfo {volume} {92}},\ \bibinfo {pages}
  {011102} (\bibinfo {year} {2004})}\BibitemShut {NoStop}%
\bibitem [{\citenamefont {Booth}\ \emph {et~al.}(2006)\citenamefont {Booth},
  \citenamefont {Brits}, \citenamefont {Gonz{\'a}lez},\ and\ \citenamefont {Van
  Den~Broeck}}]{Booth:2005ng}%
  \BibitemOpen
  \bibfield  {author} {\bibinfo {author} {\bibfnamefont {I.}~\bibnamefont
  {Booth}}, \bibinfo {author} {\bibfnamefont {L.}~\bibnamefont {Brits}},
  \bibinfo {author} {\bibfnamefont {J.~A.}\ \bibnamefont {Gonz{\'a}lez}}, \
  and\ \bibinfo {author} {\bibfnamefont {C.}~\bibnamefont {Van Den~Broeck}},\
  }\href@noop {} {\bibfield  {journal} {\bibinfo  {journal} {Classical Quantum
  Gravity}\ }\textbf {\bibinfo {volume} {23}},\ \bibinfo {pages} {413}
  (\bibinfo {year} {2006})}\BibitemShut {NoStop}%
\bibitem [{\citenamefont {Nielsen}\ and\ \citenamefont
  {Visser}(2006)}]{Nielsen:2005af}%
  \BibitemOpen
  \bibfield  {author} {\bibinfo {author} {\bibfnamefont {A.~B.}\ \bibnamefont
  {Nielsen}}\ and\ \bibinfo {author} {\bibfnamefont {M.}~\bibnamefont
  {Visser}},\ }\href {\doibase 10.1088/0264-9381/23/14/006} {\bibfield
  {journal} {\bibinfo  {journal} {Class.Quant.Grav.}\ }\textbf {\bibinfo
  {volume} {23}},\ \bibinfo {pages} {4637} (\bibinfo {year}
  {2006})}\BibitemShut {NoStop}%
\bibitem [{\citenamefont {Schnetter}\ \emph {et~al.}(2006)\citenamefont
  {Schnetter}, \citenamefont {Krishnan},\ and\ \citenamefont
  {Beyer}}]{Schnetter-Krishnan-Beyer-2006}%
  \BibitemOpen
  \bibfield  {author} {\bibinfo {author} {\bibfnamefont {E.}~\bibnamefont
  {Schnetter}}, \bibinfo {author} {\bibfnamefont {B.}~\bibnamefont {Krishnan}},
  \ and\ \bibinfo {author} {\bibfnamefont {F.}~\bibnamefont {Beyer}},\
  }\href@noop {} {\bibfield  {journal} {\bibinfo  {journal} {Phys. Rev. D}\
  }\textbf {\bibinfo {volume} {74}},\ \bibinfo {pages} {024028} (\bibinfo
  {year} {2006})}\BibitemShut {NoStop}%
\bibitem [{\citenamefont {Booth}\ and\ \citenamefont
  {Fairhurst}(2008)}]{Booth:2007wu}%
  \BibitemOpen
  \bibfield  {author} {\bibinfo {author} {\bibfnamefont {I.}~\bibnamefont
  {Booth}}\ and\ \bibinfo {author} {\bibfnamefont {S.}~\bibnamefont
  {Fairhurst}},\ }\href {\doibase 10.1103/PhysRevD.77.084005} {\bibfield
  {journal} {\bibinfo  {journal} {Phys. Rev.}\ }\textbf {\bibinfo {volume}
  {D77}},\ \bibinfo {pages} {084005} (\bibinfo {year} {2008})}\BibitemShut
  {NoStop}%
\bibitem [{\citenamefont {Jaramillo}\ \emph {et~al.}(2009)\citenamefont
  {Jaramillo}, \citenamefont {Ansorg},\ and\ \citenamefont
  {Vasset}}]{Jaramillo:2009zz}%
  \BibitemOpen
  \bibfield  {author} {\bibinfo {author} {\bibfnamefont {J.~L.}\ \bibnamefont
  {Jaramillo}}, \bibinfo {author} {\bibfnamefont {M.}~\bibnamefont {Ansorg}}, \
  and\ \bibinfo {author} {\bibfnamefont {N.}~\bibnamefont {Vasset}},\ }\href
  {\doibase 10.1063/1.3141305} {\bibfield  {journal} {\bibinfo  {journal} {AIP
  Conf. Proc.}\ }\textbf {\bibinfo {volume} {1122}},\ \bibinfo {pages} {308}
  (\bibinfo {year} {2009})}\BibitemShut {NoStop}%
\bibitem [{\citenamefont {Booth}\ and\ \citenamefont
  {Fairhurst}(2007)}]{Booth:2006bn}%
  \BibitemOpen
  \bibfield  {author} {\bibinfo {author} {\bibfnamefont {I.}~\bibnamefont
  {Booth}}\ and\ \bibinfo {author} {\bibfnamefont {S.}~\bibnamefont
  {Fairhurst}},\ }\href {\doibase 10.1103/PhysRevD.75.084019} {\bibfield
  {journal} {\bibinfo  {journal} {Phys. Rev. D}\ }\textbf {\bibinfo {volume}
  {75}},\ \bibinfo {pages} {084019} (\bibinfo {year} {2007})}\BibitemShut
  {NoStop}%
\bibitem [{\citenamefont {Ashtekar}\ and\ \citenamefont
  {Krishnan}(2002{\natexlab{a}})}]{Ashtekar02a}%
  \BibitemOpen
  \bibfield  {author} {\bibinfo {author} {\bibfnamefont {A.}~\bibnamefont
  {Ashtekar}}\ and\ \bibinfo {author} {\bibfnamefont {B.}~\bibnamefont
  {Krishnan}},\ }\href@noop {} {\bibfield  {journal} {\bibinfo  {journal}
  {Phys. Rev. Lett.}\ }\textbf {\bibinfo {volume} {89}},\ \bibinfo {pages}
  {261101} (\bibinfo {year} {2002}{\natexlab{a}})}\BibitemShut {NoStop}%
\bibitem [{\citenamefont {Hawking}(1971)}]{Hawking71a}%
  \BibitemOpen
  \bibfield  {author} {\bibinfo {author} {\bibfnamefont {S.}~\bibnamefont
  {Hawking}},\ }\href@noop {} {\bibfield  {journal} {\bibinfo  {journal} {Phys.
  Rev. Lett.}\ }\textbf {\bibinfo {volume} {26}},\ \bibinfo {pages} {1344}
  (\bibinfo {year} {1971})}\BibitemShut {NoStop}%
\bibitem [{\citenamefont {Hawking}(1972)}]{Hawking72a}%
  \BibitemOpen
  \bibfield  {author} {\bibinfo {author} {\bibfnamefont {S.~W.}\ \bibnamefont
  {Hawking}},\ }\href@noop {} {\bibfield  {journal} {\bibinfo  {journal} {Comm.
  Math. Phys.}\ }\textbf {\bibinfo {volume} {25}},\ \bibinfo {pages} {152}
  (\bibinfo {year} {1972})}\BibitemShut {NoStop}%
\bibitem [{\citenamefont {Ashtekar}\ and\ \citenamefont
  {Galloway}(2005)}]{Ashtekar05}%
  \BibitemOpen
  \bibfield  {author} {\bibinfo {author} {\bibfnamefont {A.}~\bibnamefont
  {Ashtekar}}\ and\ \bibinfo {author} {\bibfnamefont {G.}~\bibnamefont
  {Galloway}},\ }\href@noop {} {\bibfield  {journal} {\bibinfo  {journal}
  {Advances in Theoretical and Mathematical Physics}\ }\textbf {\bibinfo
  {volume} {9}},\ \bibinfo {pages} {1} (\bibinfo {year} {2005})}\BibitemShut
  {NoStop}%
\bibitem [{\citenamefont {Andersson}\ \emph {et~al.}(2005)\citenamefont
  {Andersson}, \citenamefont {Mars},\ and\ \citenamefont {Simon}}]{ams05}%
  \BibitemOpen
  \bibfield  {author} {\bibinfo {author} {\bibfnamefont {L.}~\bibnamefont
  {Andersson}}, \bibinfo {author} {\bibfnamefont {M.}~\bibnamefont {Mars}}, \
  and\ \bibinfo {author} {\bibfnamefont {W.}~\bibnamefont {Simon}},\
  }\href@noop {} {\bibfield  {journal} {\bibinfo  {journal} {Phys. Rev. Lett.}\
  }\textbf {\bibinfo {volume} {95}},\ \bibinfo {pages} {111102} (\bibinfo
  {year} {2005})}\BibitemShut {NoStop}%
\bibitem [{\citenamefont {Andersson}\ \emph {et~al.}(2008)\citenamefont
  {Andersson}, \citenamefont {Mars},\ and\ \citenamefont
  {Simon}}]{AndMarSim07}%
  \BibitemOpen
  \bibfield  {author} {\bibinfo {author} {\bibfnamefont {L.}~\bibnamefont
  {Andersson}}, \bibinfo {author} {\bibfnamefont {M.}~\bibnamefont {Mars}}, \
  and\ \bibinfo {author} {\bibfnamefont {W.}~\bibnamefont {Simon}},\
  }\href@noop {} {\bibfield  {journal} {\bibinfo  {journal} {Adv. Theor. Math.
  Phys.}\ }\textbf {\bibinfo {volume} {12}},\ \bibinfo {pages} {853} (\bibinfo
  {year} {2008})}\BibitemShut {NoStop}%
\bibitem [{\citenamefont {Ashtekar}\ and\ \citenamefont
  {Krishnan}(2002{\natexlab{b}})}]{Ashtekar-etal-2002-dynamical-horizons}%
  \BibitemOpen
  \bibfield  {author} {\bibinfo {author} {\bibfnamefont {A.}~\bibnamefont
  {Ashtekar}}\ and\ \bibinfo {author} {\bibfnamefont {B.}~\bibnamefont
  {Krishnan}},\ }\href@noop {} {\bibfield  {journal} {\bibinfo  {journal}
  {Phys. Rev. Lett.}\ }\textbf {\bibinfo {volume} {89}},\ \bibinfo {pages}
  {261101} (\bibinfo {year} {2002}{\natexlab{b}})}\BibitemShut {NoStop}%
\bibitem [{\citenamefont {Hayward}(2004{\natexlab{a}})}]{Hay04}%
  \BibitemOpen
  \bibfield  {author} {\bibinfo {author} {\bibfnamefont {S.}~\bibnamefont
  {Hayward}},\ }\href@noop {} {\bibfield  {journal} {\bibinfo  {journal} {Phys.
  Rev. Lett.}\ }\textbf {\bibinfo {volume} {93}},\ \bibinfo {pages} {251101}
  (\bibinfo {year} {2004}{\natexlab{a}})}\BibitemShut {NoStop}%
\bibitem [{\citenamefont {Hayward}(2004{\natexlab{b}})}]{Hayward04}%
  \BibitemOpen
  \bibfield  {author} {\bibinfo {author} {\bibfnamefont {S.~A.}\ \bibnamefont
  {Hayward}},\ }\href@noop {} {\bibfield  {journal} {\bibinfo  {journal} {Phys.
  Rev. D}\ }\textbf {\bibinfo {volume} {70}},\ \bibinfo {pages} {104027}
  (\bibinfo {year} {2004}{\natexlab{b}})}\BibitemShut {NoStop}%
\bibitem [{\citenamefont {Jaramillo}\ \emph {et~al.}(2008)\citenamefont
  {Jaramillo}, \citenamefont {Vasset},\ and\ \citenamefont
  {Ansorg}}]{JarVasAns07}%
  \BibitemOpen
  \bibfield  {author} {\bibinfo {author} {\bibfnamefont {J.~L.}\ \bibnamefont
  {Jaramillo}}, \bibinfo {author} {\bibfnamefont {N.}~\bibnamefont {Vasset}}, \
  and\ \bibinfo {author} {\bibfnamefont {M.}~\bibnamefont {Ansorg}},\
  }\href@noop {} {\bibfield  {journal} {\bibinfo  {journal} {EAS Publications
  Series}\ }\textbf {\bibinfo {volume} {30}},\ \bibinfo {pages} {257} (\bibinfo
  {year} {2008})}\BibitemShut {NoStop}%
\bibitem [{\citenamefont {Szekeres}(1965)}]{Szekeres65}%
  \BibitemOpen
  \bibfield  {author} {\bibinfo {author} {\bibfnamefont {P.}~\bibnamefont
  {Szekeres}},\ }\href@noop {} {\bibfield  {journal} {\bibinfo  {journal} {J.
  Math. Phys}\ }\textbf {\bibinfo {volume} {6}},\ \bibinfo {pages} {1387}
  (\bibinfo {year} {1965})}\BibitemShut {NoStop}%
\bibitem [{\citenamefont {Nolan}(2004)}]{Nolan04}%
  \BibitemOpen
  \bibfield  {author} {\bibinfo {author} {\bibfnamefont {B.~C.}\ \bibnamefont
  {Nolan}},\ }\href {\doibase 10.1103/PhysRevD.70.044004} {\bibfield  {journal}
  {\bibinfo  {journal} {Phys. Rev. D}\ }\textbf {\bibinfo {volume} {70}},\
  \bibinfo {pages} {044004} (\bibinfo {year} {2004})}\BibitemShut {NoStop}%
\bibitem [{\citenamefont {Krishnan}\ \emph {et~al.}(2007)\citenamefont
  {Krishnan}, \citenamefont {Lousto},\ and\ \citenamefont
  {Zlochower}}]{Krishnan:2007pu}%
  \BibitemOpen
  \bibfield  {author} {\bibinfo {author} {\bibfnamefont {B.}~\bibnamefont
  {Krishnan}}, \bibinfo {author} {\bibfnamefont {C.~O.}\ \bibnamefont
  {Lousto}}, \ and\ \bibinfo {author} {\bibfnamefont {Y.}~\bibnamefont
  {Zlochower}},\ }\href {\doibase 10.1103/PhysRevD.76.081501} {\bibfield
  {journal} {\bibinfo  {journal} {Phys. Rev. D}\ }\textbf {\bibinfo {volume}
  {76}},\ \bibinfo {pages} {081501} (\bibinfo {year} {2007})}\BibitemShut
  {NoStop}%
\bibitem [{\citenamefont {Lovelace}\ \emph {et~al.}(2010)\citenamefont
  {Lovelace} \emph {et~al.}}]{Lovelace:2009dg}%
  \BibitemOpen
  \bibfield  {author} {\bibinfo {author} {\bibfnamefont {G.}~\bibnamefont
  {Lovelace}} \emph {et~al.},\ }\href {\doibase 10.1103/PhysRevD.82.064031}
  {\bibfield  {journal} {\bibinfo  {journal} {Phys. Rev. D}\ }\textbf {\bibinfo
  {volume} {82}},\ \bibinfo {pages} {064031} (\bibinfo {year}
  {2010})}\BibitemShut {NoStop}%
\bibitem [{\citenamefont {Wu}(2006)}]{Wu06}%
  \BibitemOpen
  \bibfield  {author} {\bibinfo {author} {\bibfnamefont {Y.-H.}\ \bibnamefont
  {Wu}},\ }\emph {\bibinfo {title} {Isolated Horizons, Dynamical Horizons and
  their quasi-local energy-momentum and flux}},\ \href@noop {} {Ph.D. thesis},\
  \bibinfo  {school} {University of Southampton}, \bibinfo {address}
  {Southampton, U.K.} (\bibinfo {year} {2006})\BibitemShut {NoStop}%
\bibitem [{\citenamefont {Wu}\ and\ \citenamefont {Wang}(2009)}]{Wu:2009dm}%
  \BibitemOpen
  \bibfield  {author} {\bibinfo {author} {\bibfnamefont {Y.-H.}\ \bibnamefont
  {Wu}}\ and\ \bibinfo {author} {\bibfnamefont {C.-H.}\ \bibnamefont {Wang}},\
  }\href {\doibase 10.1103/PhysRevD.80.063002} {\bibfield  {journal} {\bibinfo
  {journal} {Phys. Rev. D}\ }\textbf {\bibinfo {volume} {80}},\ \bibinfo
  {pages} {063002} (\bibinfo {year} {2009})}\BibitemShut {NoStop}%
\bibitem [{\citenamefont {Wu}\ and\ \citenamefont {Wang}(2011)}]{Wu:2010eu}%
  \BibitemOpen
  \bibfield  {author} {\bibinfo {author} {\bibfnamefont {Y.-H.}\ \bibnamefont
  {Wu}}\ and\ \bibinfo {author} {\bibfnamefont {C.-H.}\ \bibnamefont {Wang}},\
  }\href {\doibase 10.1103/PhysRevD.83.084044} {\bibfield  {journal} {\bibinfo
  {journal} {Phys. Rev.}\ }\textbf {\bibinfo {volume} {D83}},\ \bibinfo {pages}
  {084044} (\bibinfo {year} {2011})}\BibitemShut {NoStop}%
\bibitem [{\citenamefont {Hawking}\ and\ \citenamefont
  {Hartle}(1972)}]{Hawking:1972hy}%
  \BibitemOpen
  \bibfield  {author} {\bibinfo {author} {\bibfnamefont {S.~W.}\ \bibnamefont
  {Hawking}}\ and\ \bibinfo {author} {\bibfnamefont {J.~B.}\ \bibnamefont
  {Hartle}},\ }\href {\doibase 10.1007/BF01645515} {\bibfield  {journal}
  {\bibinfo  {journal} {Commun. Math. Phys.}\ }\textbf {\bibinfo {volume}
  {27}},\ \bibinfo {pages} {283} (\bibinfo {year} {1972})}\BibitemShut
  {NoStop}%
\bibitem [{\citenamefont {Hartle}(1973)}]{Hartle:1973zz}%
  \BibitemOpen
  \bibfield  {author} {\bibinfo {author} {\bibfnamefont {J.~B.}\ \bibnamefont
  {Hartle}},\ }\href {\doibase 10.1103/PhysRevD.8.1010} {\bibfield  {journal}
  {\bibinfo  {journal} {Phys. Rev. D}\ }\textbf {\bibinfo {volume} {8}},\
  \bibinfo {pages} {1010} (\bibinfo {year} {1973})}\BibitemShut {NoStop}%
\bibitem [{\citenamefont {Hartle}(1974)}]{Hartle:1974gy}%
  \BibitemOpen
  \bibfield  {author} {\bibinfo {author} {\bibfnamefont {J.~B.}\ \bibnamefont
  {Hartle}},\ }\href {\doibase 10.1103/PhysRevD.9.2749} {\bibfield  {journal}
  {\bibinfo  {journal} {Phys. Rev. D}\ }\textbf {\bibinfo {volume} {9}},\
  \bibinfo {pages} {2749} (\bibinfo {year} {1974})}\BibitemShut {NoStop}%
\bibitem [{\citenamefont {Damour}(1979)}]{Damour79}%
  \BibitemOpen
  \bibfield  {author} {\bibinfo {author} {\bibfnamefont {T.}~\bibnamefont
  {Damour}},\ }\href@noop {} {\emph {\bibinfo {title} {Quelques propri\'et\'es
  m\'ecaniques, \'electromagn\'etiques, thermodynamiques et quantiques des
  trous noirs}}}\ (\bibinfo  {publisher} {Th\`ese de doctorat d’\'Etat,
  Universit\'e Paris 6},\ \bibinfo {address} {France},\ \bibinfo {year}
  {1979})\BibitemShut {NoStop}%
\bibitem [{\citenamefont {Damour}(1982)}]{Damour82}%
  \BibitemOpen
  \bibfield  {author} {\bibinfo {author} {\bibfnamefont {T.}~\bibnamefont
  {Damour}},\ }in\ \href@noop {} {\emph {\bibinfo {booktitle} {Proceedings of
  the Second Marcell Grossman Meeting on General Relativity}}},\ \bibinfo
  {editor} {edited by\ \bibinfo {editor} {\bibfnamefont {R.}~\bibnamefont
  {Ruffini}}}\ (\bibinfo  {publisher} {North Holland},\ \bibinfo {year}
  {1982})\ p.\ \bibinfo {pages} {587}\BibitemShut {NoStop}%
\bibitem [{\citenamefont {Price}\ and\ \citenamefont {Thorne}(1986)}]{Price86}%
  \BibitemOpen
  \bibfield  {author} {\bibinfo {author} {\bibfnamefont {R.~H.}\ \bibnamefont
  {Price}}\ and\ \bibinfo {author} {\bibfnamefont {K.~S.}\ \bibnamefont
  {Thorne}},\ }\href@noop {} {\bibfield  {journal} {\bibinfo  {journal} {Phys.
  Rev. D}\ }\textbf {\bibinfo {volume} {33}},\ \bibinfo {pages} {915} (\bibinfo
  {year} {1986})}\BibitemShut {NoStop}%
\bibitem [{\citenamefont {Crowley}\ \emph {et~al.}(1986)\citenamefont
  {Crowley}, \citenamefont {Macdonald}, \citenamefont {Price}, \citenamefont
  {Redmount}, \citenamefont {Suen}, \citenamefont {Thorne},\ and\ \citenamefont
  {Zhang}}]{Price86a}%
  \BibitemOpen
  \bibfield  {author} {\bibinfo {author} {\bibfnamefont {R.}~\bibnamefont
  {Crowley}}, \bibinfo {author} {\bibfnamefont {D.}~\bibnamefont {Macdonald}},
  \bibinfo {author} {\bibfnamefont {R.}~\bibnamefont {Price}}, \bibinfo
  {author} {\bibfnamefont {I.}~\bibnamefont {Redmount}}, \bibinfo {author}
  {\bibnamefont {Suen}}, \bibinfo {author} {\bibfnamefont {K.}~\bibnamefont
  {Thorne}}, \ and\ \bibinfo {author} {\bibfnamefont {X.-H.}\ \bibnamefont
  {Zhang}},\ }\href@noop {} {\emph {\bibinfo {title} {Black Holes: The Membrane
  Paradigm}}}\ (\bibinfo  {publisher} {Yale University Press},\ \bibinfo {year}
  {1986})\BibitemShut {NoStop}%
\bibitem [{\citenamefont {Straumann}(1997)}]{Straumann:1997tg}%
  \BibitemOpen
  \bibfield  {author} {\bibinfo {author} {\bibfnamefont {N.}~\bibnamefont
  {Straumann}},\ }\href@noop {} {\bibfield  {journal} {\bibinfo  {journal}
  {arXiv:9711276 [astro-ph]}\ } (\bibinfo {year} {1997})}\BibitemShut {NoStop}%
\bibitem [{\citenamefont {Damour}\ and\ \citenamefont
  {Lilley}(2008)}]{Damour:2008ji}%
  \BibitemOpen
  \bibfield  {author} {\bibinfo {author} {\bibfnamefont {T.}~\bibnamefont
  {Damour}}\ and\ \bibinfo {author} {\bibfnamefont {M.}~\bibnamefont
  {Lilley}},\ }\href@noop {} {\bibfield  {journal} {\bibinfo  {journal}
  {arXiv:0802.4169 [hep-th]}\ } (\bibinfo {year} {2008})}\BibitemShut {NoStop}%
\bibitem [{\citenamefont {Gourgoulhon}\ and\ \citenamefont
  {Jaramillo}(2006{\natexlab{a}})}]{Gourgoulhon-Jaramillo-Review}%
  \BibitemOpen
  \bibfield  {author} {\bibinfo {author} {\bibfnamefont {E.}~\bibnamefont
  {Gourgoulhon}}\ and\ \bibinfo {author} {\bibfnamefont {J.~L.}\ \bibnamefont
  {Jaramillo}},\ }\href {\doibase 10.1016/j.physrep.2005.10.00} {\bibfield
  {journal} {\bibinfo  {journal} {Physics Reports}\ }\textbf {\bibinfo {volume}
  {423}},\ \bibinfo {pages} {159} (\bibinfo {year}
  {2006}{\natexlab{a}})}\BibitemShut {NoStop}%
\bibitem [{\citenamefont {Gourgoulhon}(2005)}]{Gourgoulhon:2005ch}%
  \BibitemOpen
  \bibfield  {author} {\bibinfo {author} {\bibfnamefont {E.}~\bibnamefont
  {Gourgoulhon}},\ }\href@noop {} {\bibfield  {journal} {\bibinfo  {journal}
  {Phys. Rev. D}\ }\textbf {\bibinfo {volume} {72}},\ \bibinfo {pages} {104007}
  (\bibinfo {year} {2005})}\BibitemShut {NoStop}%
\bibitem [{\citenamefont {Gourgoulhon}\ and\ \citenamefont
  {Jaramillo}(2006{\natexlab{b}})}]{Gourgoulhon:2006uc}%
  \BibitemOpen
  \bibfield  {author} {\bibinfo {author} {\bibfnamefont {E.}~\bibnamefont
  {Gourgoulhon}}\ and\ \bibinfo {author} {\bibfnamefont {J.~L.}\ \bibnamefont
  {Jaramillo}},\ }\href {\doibase 10.1103/PhysRevD.74.087502} {\bibfield
  {journal} {\bibinfo  {journal} {Phys. Rev. D}\ }\textbf {\bibinfo {volume}
  {74}},\ \bibinfo {pages} {087502} (\bibinfo {year}
  {2006}{\natexlab{b}})}\BibitemShut {NoStop}%
\bibitem [{\citenamefont {Gourgoulhon}\ and\ \citenamefont
  {Jaramillo}(2008)}]{Gourgoulhon:2008pu}%
  \BibitemOpen
  \bibfield  {author} {\bibinfo {author} {\bibfnamefont {E.}~\bibnamefont
  {Gourgoulhon}}\ and\ \bibinfo {author} {\bibfnamefont {J.~L.}\ \bibnamefont
  {Jaramillo}},\ }\href {\doibase 10.1016/j.newar.2008.03.026} {\bibfield
  {journal} {\bibinfo  {journal} {New Astron. Rev.}\ }\textbf {\bibinfo
  {volume} {51}},\ \bibinfo {pages} {791} (\bibinfo {year} {2008})}\BibitemShut
  {NoStop}%
\bibitem [{\citenamefont {Padmanabhan}(2011)}]{Padma11}%
  \BibitemOpen
  \bibfield  {author} {\bibinfo {author} {\bibfnamefont {T.}~\bibnamefont
  {Padmanabhan}},\ }\href {\doibase 10.1103/PhysRevD.83.044048} {\bibfield
  {journal} {\bibinfo  {journal} {Phys. Rev. D}\ }\textbf {\bibinfo {volume}
  {83}},\ \bibinfo {pages} {044048} (\bibinfo {year} {2011})}\BibitemShut
  {NoStop}%
\bibitem [{\citenamefont {Bhattacharyya}\ \emph
  {et~al.}(2008{\natexlab{a}})\citenamefont {Bhattacharyya}, \citenamefont
  {Minwalla}, \citenamefont {Hubeny},\ and\ \citenamefont
  {Rangamani}}]{BhaMinHub08}%
  \BibitemOpen
  \bibfield  {author} {\bibinfo {author} {\bibfnamefont {S.}~\bibnamefont
  {Bhattacharyya}}, \bibinfo {author} {\bibfnamefont {S.}~\bibnamefont
  {Minwalla}}, \bibinfo {author} {\bibfnamefont {V.~E.}\ \bibnamefont
  {Hubeny}}, \ and\ \bibinfo {author} {\bibfnamefont {M.}~\bibnamefont
  {Rangamani}},\ }\href {http://stacks.iop.org/1126-6708/2008/i=02/a=045}
  {\bibfield  {journal} {\bibinfo  {journal} {Journal of High Energy Physics}\
  }\textbf {\bibinfo {volume} {2008}},\ \bibinfo {pages} {045} (\bibinfo {year}
  {2008}{\natexlab{a}})}\BibitemShut {NoStop}%
\bibitem [{\citenamefont {Bhattacharyya}\ \emph
  {et~al.}(2008{\natexlab{b}})\citenamefont {Bhattacharyya}, \citenamefont
  {Hubeny}, \citenamefont {Loganayagam}, \citenamefont {Mandal}, \citenamefont
  {Minwalla}, \citenamefont {Morita}, \citenamefont {Rangamani},\ and\
  \citenamefont {Reall}}]{BhaHubLog08}%
  \BibitemOpen
  \bibfield  {author} {\bibinfo {author} {\bibfnamefont {S.}~\bibnamefont
  {Bhattacharyya}}, \bibinfo {author} {\bibfnamefont {V.~E.}\ \bibnamefont
  {Hubeny}}, \bibinfo {author} {\bibfnamefont {R.}~\bibnamefont {Loganayagam}},
  \bibinfo {author} {\bibfnamefont {G.}~\bibnamefont {Mandal}}, \bibinfo
  {author} {\bibfnamefont {S.}~\bibnamefont {Minwalla}}, \bibinfo {author}
  {\bibfnamefont {T.}~\bibnamefont {Morita}}, \bibinfo {author} {\bibfnamefont
  {M.}~\bibnamefont {Rangamani}}, \ and\ \bibinfo {author} {\bibfnamefont
  {H.~S.}\ \bibnamefont {Reall}},\ }\href
  {http://stacks.iop.org/1126-6708/2008/i=06/a=055} {\bibfield  {journal}
  {\bibinfo  {journal} {Journal of High Energy Physics}\ }\textbf {\bibinfo
  {volume} {2008}},\ \bibinfo {pages} {055} (\bibinfo {year}
  {2008}{\natexlab{b}})}\BibitemShut {NoStop}%
\bibitem [{\citenamefont {Booth}\ \emph
  {et~al.}(2011{\natexlab{a}})\citenamefont {Booth}, \citenamefont {Heller},\
  and\ \citenamefont {Spalinski}}]{Booth:2010kr}%
  \BibitemOpen
  \bibfield  {author} {\bibinfo {author} {\bibfnamefont {I.}~\bibnamefont
  {Booth}}, \bibinfo {author} {\bibfnamefont {M.~P.}\ \bibnamefont {Heller}}, \
  and\ \bibinfo {author} {\bibfnamefont {M.}~\bibnamefont {Spalinski}},\ }\href
  {\doibase 10.1103/PhysRevD.83.061901} {\bibfield  {journal} {\bibinfo
  {journal} {Phys. Rev.}\ }\textbf {\bibinfo {volume} {D83}},\ \bibinfo {pages}
  {061901} (\bibinfo {year} {2011}{\natexlab{a}})}\BibitemShut {NoStop}%
\bibitem [{\citenamefont {Booth}\ \emph
  {et~al.}(2011{\natexlab{b}})\citenamefont {Booth}, \citenamefont {Heller},
  \citenamefont {Plewa},\ and\ \citenamefont {Spali\ifmmode~\acute{n}\else
  \'{n}\fi{}ski}}]{BooHelPle11}%
  \BibitemOpen
  \bibfield  {author} {\bibinfo {author} {\bibfnamefont {I.}~\bibnamefont
  {Booth}}, \bibinfo {author} {\bibfnamefont {M.~P.}\ \bibnamefont {Heller}},
  \bibinfo {author} {\bibfnamefont {G.}~\bibnamefont {Plewa}}, \ and\ \bibinfo
  {author} {\bibfnamefont {M.}~\bibnamefont {Spali\ifmmode~\acute{n}\else
  \'{n}\fi{}ski}},\ }\href {\doibase 10.1103/PhysRevD.83.106005} {\bibfield
  {journal} {\bibinfo  {journal} {Phys. Rev. D}\ }\textbf {\bibinfo {volume}
  {83}},\ \bibinfo {pages} {106005} (\bibinfo {year}
  {2011}{\natexlab{b}})}\BibitemShut {NoStop}%
\bibitem [{\citenamefont {Eling}\ and\ \citenamefont
  {Oz}(2010)}]{Eling:2009sj}%
  \BibitemOpen
  \bibfield  {author} {\bibinfo {author} {\bibfnamefont {C.}~\bibnamefont
  {Eling}}\ and\ \bibinfo {author} {\bibfnamefont {Y.}~\bibnamefont {Oz}},\
  }\href {\doibase 10.1007/JHEP02(2010)069} {\bibfield  {journal} {\bibinfo
  {journal} {JHEP}\ }\textbf {\bibinfo {volume} {1002}},\ \bibinfo {pages}
  {069} (\bibinfo {year} {2010})}\BibitemShut {NoStop}%
\bibitem [{\citenamefont {Owen}\ \emph {et~al.}(2011)\citenamefont {Owen},
  \citenamefont {Brink}, \citenamefont {Chen}, \citenamefont {Kaplan},
  \citenamefont {Lovelace}, \citenamefont {Matthews}, \citenamefont {Nichols},
  \citenamefont {Scheel}, \citenamefont {Zhang}, \citenamefont {Zimmerman},\
  and\ \citenamefont {Thorne}}]{OweBriChe11}%
  \BibitemOpen
  \bibfield  {author} {\bibinfo {author} {\bibfnamefont {R.}~\bibnamefont
  {Owen}}, \bibinfo {author} {\bibfnamefont {J.}~\bibnamefont {Brink}},
  \bibinfo {author} {\bibfnamefont {Y.}~\bibnamefont {Chen}}, \bibinfo {author}
  {\bibfnamefont {J.~D.}\ \bibnamefont {Kaplan}}, \bibinfo {author}
  {\bibfnamefont {G.}~\bibnamefont {Lovelace}}, \bibinfo {author}
  {\bibfnamefont {K.~D.}\ \bibnamefont {Matthews}}, \bibinfo {author}
  {\bibfnamefont {D.~A.}\ \bibnamefont {Nichols}}, \bibinfo {author}
  {\bibfnamefont {M.~A.}\ \bibnamefont {Scheel}}, \bibinfo {author}
  {\bibfnamefont {F.}~\bibnamefont {Zhang}}, \bibinfo {author} {\bibfnamefont
  {A.}~\bibnamefont {Zimmerman}}, \ and\ \bibinfo {author} {\bibfnamefont
  {K.~S.}\ \bibnamefont {Thorne}},\ }\href {\doibase
  10.1103/PhysRevLett.106.151101} {\bibfield  {journal} {\bibinfo  {journal}
  {Phys. Rev. Lett.}\ }\textbf {\bibinfo {volume} {106}},\ \bibinfo {pages}
  {151101} (\bibinfo {year} {2011})}\BibitemShut {NoStop}%
\bibitem [{\citenamefont {Nichols}\ \emph {et~al.}(2011)\citenamefont
  {Nichols}, \citenamefont {Owen}, \citenamefont {Zhang}, \citenamefont
  {Zimmerman}, \citenamefont {Brink}, \citenamefont {Chen}, \citenamefont
  {Kaplan}, \citenamefont {Lovelace}, \citenamefont {Matthews}, \citenamefont
  {Scheel},\ and\ \citenamefont {Thorne}}]{Nichols2011}%
  \BibitemOpen
  \bibfield  {author} {\bibinfo {author} {\bibfnamefont {D.~A.}\ \bibnamefont
  {Nichols}}, \bibinfo {author} {\bibfnamefont {R.}~\bibnamefont {Owen}},
  \bibinfo {author} {\bibfnamefont {F.}~\bibnamefont {Zhang}}, \bibinfo
  {author} {\bibfnamefont {A.}~\bibnamefont {Zimmerman}}, \bibinfo {author}
  {\bibfnamefont {J.}~\bibnamefont {Brink}}, \bibinfo {author} {\bibfnamefont
  {Y.}~\bibnamefont {Chen}}, \bibinfo {author} {\bibfnamefont {J.~D.}\
  \bibnamefont {Kaplan}}, \bibinfo {author} {\bibfnamefont {G.}~\bibnamefont
  {Lovelace}}, \bibinfo {author} {\bibfnamefont {K.~D.}\ \bibnamefont
  {Matthews}}, \bibinfo {author} {\bibfnamefont {M.~A.}\ \bibnamefont
  {Scheel}}, \ and\ \bibinfo {author} {\bibfnamefont {K.~S.}\ \bibnamefont
  {Thorne}},\ }\href {\doibase 10.1103/PhysRevD.84.124014} {\bibfield
  {journal} {\bibinfo  {journal} {Phys. Rev. D}\ }\textbf {\bibinfo {volume}
  {84}},\ \bibinfo {pages} {124014} (\bibinfo {year} {2011})}\BibitemShut
  {NoStop}%
\bibitem [{\citenamefont {Cao}(2011)}]{Cao:2010vj}%
  \BibitemOpen
  \bibfield  {author} {\bibinfo {author} {\bibfnamefont {L.-M.}\ \bibnamefont
  {Cao}},\ }\href {\doibase 10.1007/JHEP03(2011)112} {\bibfield  {journal}
  {\bibinfo  {journal} {JHEP}\ }\textbf {\bibinfo {volume} {03}},\ \bibinfo
  {pages} {112} (\bibinfo {year} {2011})}\BibitemShut {NoStop}%
\bibitem [{\citenamefont {Szil{\'a}gyi}\ \emph {et~al.}(2007)\citenamefont
  {Szil{\'a}gyi}, \citenamefont {Pollney}, \citenamefont {Rezzolla},
  \citenamefont {Thornburg},\ and\ \citenamefont {Winicour}}]{Szilagyi:2006qy}%
  \BibitemOpen
  \bibfield  {author} {\bibinfo {author} {\bibfnamefont {B.}~\bibnamefont
  {Szil{\'a}gyi}}, \bibinfo {author} {\bibfnamefont {D.}~\bibnamefont
  {Pollney}}, \bibinfo {author} {\bibfnamefont {L.}~\bibnamefont {Rezzolla}},
  \bibinfo {author} {\bibfnamefont {J.}~\bibnamefont {Thornburg}}, \ and\
  \bibinfo {author} {\bibfnamefont {J.}~\bibnamefont {Winicour}},\ }\href@noop
  {} {\bibfield  {journal} {\bibinfo  {journal} {Class. Quant. Grav.}\ }\textbf
  {\bibinfo {volume} {24}},\ \bibinfo {pages} {S275} (\bibinfo {year}
  {2007})}\BibitemShut {NoStop}%
\end{thebibliography}%

\end{document}